\documentclass[prd,aps,showpacs,nofootinbib,preprint,eqsecnum]{revtex4}
\usepackage{amsmath}
\usepackage{amssymb}
\usepackage{amsfonts}
\usepackage{graphicx,bm,color}
\usepackage{dcolumn}
\usepackage{color,amsxtra}
\usepackage{epsf}
\usepackage{enumerate}
\usepackage{hhline}
\usepackage{array}
\usepackage{tabularx}
\newcommand{\be}{\begin{equation}}
\newcommand{\ee}{\end{equation}}
\newcommand{\bea}{\begin{eqnarray}}
\newcommand{\eea}{\end{eqnarray}}
\newcommand{\beaa}{\begin{eqnarray*}}
\newcommand{\eeaa}{\end{eqnarray*}}

\newcommand{\nn}{\nonumber \\}
\newcommand{\e}{\mathrm{e}}

\newcommand{\Eqn}[1]{&\hspace{-0.2em}#1\hspace{-0.2em}&}
\newcommand{\abs}[1]{\vert{#1}\vert}

\begin{document}


\title{Screening of cosmological constant for De Sitter Universe in non-local 
gravity, phantom-divide crossing and finite-time future singularities
}

\author{Kazuharu Bamba$^{1,}$\footnote{
E-mail address: bamba@kmi.nagoya-u.ac.jp},
Shin'ichi Nojiri$^{1, 2,}$\footnote{E-mail address:
nojiri@phys.nagoya-u.ac.jp}, 
Sergei D. Odintsov$^{3, 4,}$\footnote{
E-mail address: odintsov@ieec.uab.es} 
and 
Misao Sasaki$^{5, 6,}$\footnote{
E-mail address: misao@yukawa.kyoto-u.ac.jp}}
\affiliation{
$^1$Kobayashi-Maskawa Institute for the Origin of Particles and the
Universe,
Nagoya University, Nagoya 464-8602, Japan\\
$^2$Department of Physics, Nagoya University, Nagoya 464-8602, Japan\\
$^3$Instituci\`{o} Catalana de Recerca i Estudis Avan\c{c}ats (ICREA)
and Institut de Ciencies de l'Espai (IEEC-CSIC),
Campus UAB, Facultat de Ciencies, Torre C5-Par-2a pl, E-08193 Bellaterra
(Barcelona), Spain\\
$^4$Tomsk State Pedagogical University, Tomsk, Russia\\
$^5$Yukawa Institute for Theoretical Physics, 
Kyoto University, Kyoto 606-8502, Japan\\
$^6$Korea Institute for Advanced Study
207-43 Cheongnyangni 2-dong, Dongdaemun-gu, 
Seoul 130-722, Republic of Korea
}


\begin{abstract}

We investigate de Sitter solutions in non-local gravity as well as 
in non-local gravity with Lagrange constraint multiplier. 
We examine a condition to avoid a ghost and discuss a screening scenario 
for a cosmological constant in de Sitter solutions. 
Furthermore, we explicitly demonstrate that 
three types of the finite-time future singularities 
can occur in non-local gravity and explore their properties. 
In addition, we evaluate the effective equation of state for the universe 
and show that the late-time accelerating universe may be effectively the 
quintessence, cosmological constant or phantom-like phases. 
In particular, it is found that there is a case in which a crossing of 
the phantom divide from the non-phantom (quintessence) phase to the phantom 
one can be realized when a finite-time future singularity occurs. 
Moreover, it is demonstrated that the addition of an $R^2$ term 
can cure the finite-time future singularities in non-local gravity. 
It is also suggested that in the framework of non-local gravity, 
adding an $R^2$ term leads to possible unification of the 
early-time inflation with the late-time cosmic acceleration.

\end{abstract}

\pacs{
04.50.Kd, 95.36.+x, 98.80.-k
}
\preprint{YITP-11-46}

\maketitle

\section{Introduction}

It has been suggested that the current expansion of the universe 
is accelerating 
by recent cosmological observations such as 
Supernovae Ia (SNe Ia)~\cite{SN1}, 
cosmic microwave background (CMB) radiation~\cite{WMAP, Komatsu:2010fb}, 
large scale structure (LSS)~\cite{LSS},  
baryon acoustic oscillations (BAO)~\cite{Eisenstein:2005su}, 
and weak lensing~\cite{Jain:2003tba}. 
There are two representative approaches to account for 
the late time cosmic acceleration. 
One is the introduction of cosmological constant dark energy 
in the framework of general relativity 
(for a review, see, e.g.,~\cite{Weinberg:1988cp}). 
The other is the modification of gravity, 
for example, $F(R)$ gravity, 
where $F(R)$ is an arbitrary function of the scalar curvature $R$ 
(for recent reviews on $F(R)$ gravity, 
see, e.g.,~\cite{Review-Nojiri-Odintsov, Book-Capozziello-Faraoni}). 

As a possible 
modification of gravity, 
non-local gravity produced by quantum effects 
has been proposed in Ref.~\cite{Deser:2007jk}. 
A modification of non-local gravity with an $F(R)$ term 
has been studied in order to realize a unified scenario of 
the early-time inflation with the late-time cosmic 
acceleration~\cite{Nojiri:2007uq}. 
Moreover, 
there was a proposal on the solution of the cosmological constant 
problem by non-local modification of gravity~\cite{ArkaniHamed:2002fu}. 
Recently, an explicit mechanism to screen a cosmological constant in non-local 
gravity has been discussed in Ref.~\cite{Nojiri:2010pw}. 
It is also significant to mention that 
there exists the problem of ghosts in non-local gravity~\cite{Nojiri:2010pw}.
Non-local gravity has been extensively examined 
in the literature~\cite{Non-local-gravity-Refs, 
Koivisto:2008xfa, Koivisto:2008dh, Elizalde:2011su}. 

Furthermore, 
it is known that 
so-called matter instability~\cite{Dolgov:2003px} 
occurs in $F(R)$ gravity. 
This implies that the curvature inside matter sphere 
becomes very large and hence the curvature singularity 
could appear. 
The problem of singularity in 
the relativistic star formation process was discussed in Ref.~\cite{maeda}. 
In addition, the instability in $F(R)$ gravity has recently been 
studied for a gravitating system with a time dependent mass density 
such as astronomical massive objects in Ref.~\cite{Arbuzova:2010iu}. 
The generation mechanism of 
the time-dependent matter instability in the star collapse 
has also been investigated in Ref.~\cite{Bamba:2011sm}. 
Note that a scenario to cure the star singularity via adding an $R^2$ term
 or some other powers of the scalar curvature has been developed in 
Refs.~\cite{Review-Nojiri-Odintsov, Bamba:2011sm}.
On the other hand, 
the stability of cosmological perturbations 
in non-local gravity has been explored in Ref.~\cite{Koivisto:2008dh}. 
It is important to examine whether 
there exists the curvature singularity in non-local gravity. 
In what follows, the curvature singularity is called 
``the finite-time future singularities'' throughout this paper.

In this paper, 
we explore de Sitter solutions not only in non-local gravity 
but also in non-local gravity with Lagrange constraint multiplier. 
We also study a condition to avoid a ghost 
and present a screening scenario for a cosmological constant 
in de Sitter solutions. 
We reconfirm that the problem of ghosts persists in these non-local models. 
This is an important result of this work. 
In addition, we explicitly show that 
three types of the finite-time future singularities 
can occur in non-local gravity and examine their properties. 
Furthermore, 
we investigate the behavior of the effective equation of state for 
the universe when the finite-time future singularities occur
and show that the late-time accelerating universe may be effectively the 
quintessence, cosmological constant or phantom-like phases. 
In particular, we demonstrate that there is a case with realizing 
a crossing of the phantom divide 
from the non-phantom (quintessence) 
phase to the phantom one 
in the limit of the appearance of 
a finite-time future singularity. 
Moreover, we consider the cosmological consequences of adding an $R^2$ term 
and demonstrate that the addition of an $R^2$ term 
can remove the finite-time future singularities in non-local gravity. 
We use units of $k_\mathrm{B} = c = \hbar = 1$ and denote the
gravitational constant $8 \pi G$ by
${\kappa}^2 \equiv 8\pi/{M_{\mathrm{Pl}}}^2$
with the Planck mass of $M_{\mathrm{Pl}} = G^{-1/2} = 1.2 \times 10^{19}$GeV.

The paper is organized as follows. 
In Sec.\ II, we review non-local gravity and study 
a de Sitter solution in non-local gravity. 
We examine a condition to avoid a ghost 
and discuss a screening scenario for a cosmological constant 
in the de Sitter solution. 
In Sec.\ III, 
we also investigate a de Sitter solution 
in non-local gravity with Lagrange constraint multiplier. 
In Sec.\ IV, we demonstrate that 
three types of the finite-time future singularities 
can occur in non-local gravity and explore their properties. 
In addition, 
we consider the behavior of the effective equation of state for the universe 
when the finite-time future singularities occur in Sec.\ V. 
Finally, conclusions are given in Sec.\ VI.

\section{de Sitter solution in non-local gravity}

\subsection{de Sitter solution}

In the previous work~\cite{Nojiri:2010pw}, 
it has been found that there exists a flat space-time solution even in the 
presence of the cosmological constant in non-local gravity, and 
a possible mechanism to screen the cosmological constant has been discussed. 
The new point in this paper is 
it is shown that there could exist a de Sitter solution and 
a effective screening scenario in the de Sitter solution is considered. 

The starting action of non-local gravity is given by
\be
\label{nl1}
S=\int d^4 x \sqrt{-g}\left\{
\frac{1}{2\kappa^2}\left[ R\left(1 + f(\Box^{-1}R )\right) -2 \Lambda \right]
+ \mathcal{L}_\mathrm{matter} \left(Q; g\right)
\right\}
\, .
\ee
Here, $g$ is the determinant of the metric tensor $g_{\mu\nu}$, 
$f$ is some function, $\Box \equiv g^{\mu \nu} {\nabla}_{\mu} {\nabla}_{\nu}$ 
with ${\nabla}_{\mu}$ being the covariant derivative 
is the covariant d'Almbertian for a scalar field, 
$\Lambda$ is a cosmological constant, 
and 
$\mathcal{L}_\mathrm{matter} \left(Q; g\right)$ is the matter Lagrangian, 
where $Q$ stands for the matter fields. 
We regard the action (\ref{nl1}) as a low energy effective action including quantum corrections. 
Then in the following, we treat this model only classically. 
We should also note that the action should not be general action including all the terms allowed 
by symmetry. This is mainly due to the technical reason since we cannot work with such a complicated  
action. 

We also remark that 
in the action (\ref{nl1}), the ultraviolet divergences 
would become worse, and at large momenta the
non-local term will contribute a large correction. 
We regard, however, that the model could be a low energy effective action 
including quantum 
corrections and hence this model should be treated only classically and 
we do not include further quantum corrections. We do not always consider 
the relation with string theory, either.  

The above action in Eq.~(\ref{nl1}) 
can be rewritten by introducing two scalar fields $\eta$ 
and $\xi$ in the following form: 
\bea
\label{nl2}
S \Eqn{=}
\int d^4 x \sqrt{-g}\left\{
\frac{1}{2\kappa^2}\left[R\left(1 + f(\eta)\right) 
+ \xi\left(\Box\eta - R\right) - 2 \Lambda \right]
+ \mathcal{L}_\mathrm{matter}  \right\} \nn 
\Eqn{=}
\int d^4 x \sqrt{-g}\left\{
\frac{1}{2\kappa^2}\left[R\left(1 + f(\eta)\right) 
 - \partial_\mu \xi \partial^\mu \eta - \xi R - 2 \Lambda \right]
+ \mathcal{L}_\mathrm{matter} 
\right\}
\, .
\eea
%
%

We take the flat Friedmann-Lema\^{i}tre-Robertson-Walker (FLRW) metric 
\be
\label{nl6}
ds^2 = - dt^2 + a^2(t) \sum_{i=1,2,3}\left(dx^i\right)^2\, . 
\ee
We consider the case in which 
the scalar fields $\eta$ and $\xi$ only depend on time. 
In this background, 
the 
gravitational field equations are given by 
\bea
\label{nl7a}
\hspace{-10mm}
0 \Eqn{=} 
- 3 H^2\left(1 + f(\eta) - \xi\right) + \frac{1}{2}\dot\xi \dot\eta 
 - 3H\left(f'(\eta)\dot\eta - \dot\xi\right) + \Lambda 
+ \kappa^2 \rho_{\mathrm{m}}\, ,\\
\label{nl7b}
\hspace{-10mm}
0 \Eqn{=} 
\left(2\dot H + 3H^2\right) \left(1 + f(\eta) - \xi\right) 
+ \frac{1}{2}\dot\xi \dot\eta 
+ \left(\frac{d^2}{dt^2} + 2H \frac{d}{dt} \right) \left( f(\eta) - 
\xi \right) - \Lambda + \kappa^2 P_{\mathrm{m}}\, , 
\eea
where $H=\dot{a}/a$ is the Hubble parameter, 
the dot denotes the time derivative of $\partial/\partial t$, and 
$\rho_{\mathrm{m}}$ and 
$P_{\mathrm{m}}$ 
are the energy density and pressure of matter, 
respectively. 
For a perfect fluid of matter, we have 
$T_{\mathrm{matter}\, 0 0} = \rho_{\mathrm{m}}$ and 
$T_{\mathrm{matter}\, i j} = P_{\mathrm{m}} \delta_{i j}$. 
Furthermore, 
the equations of motion for the scalar fields $\eta$ and $\xi$ 
are given by
\bea
\label{nl8a}
0 \Eqn{=} 
\ddot \eta + 3H \dot \eta + 6 \dot H + 12 H^2 \, , \\
\label{nl8b}
0 \Eqn{=} 
\ddot \xi + 3H \dot \xi - \left( 6 \dot H + 12 H^2\right)f'(\eta) \, , 
\eea
where we have used $R = 6\dot{H} + 12H^2$. 

We 
note that in Ref.~\cite{Deser:2007jk}, 
the following points are stated. 
The reason why the non-local contribution $\Box^{-1}$ is taken is that 
it is a simplest choice for the 
the inverse of some differential operator to
provide the required time lag between the transition from 
radiation dominance to matter dominance at the radiation-matter 
equality time $t_\mathrm{eq} \sim 10^5$ years. 
Much larger values can be obtained through other operators,
for example, the Paneitz operator arising in the 
context of conformal anomalies~\cite{Conformal-anomalies}, 
which is given by 
$\left( 1/\sqrt{-g} \right) \triangle_{P}$. 
One gets about $10^6$ from the dimensionless combination of
the inverse of this operator acting on $R^2$. 


We now assume a de Sitter solution $H=H_0$, where $H_0$ is a constant. 
Then, Eq.~(\ref{nl8a}) can be solved 
as 
\be
\label{NLdS1}
\eta= - 4H_0 t - \eta_0 \e^{-3H_0 t} + \eta_1\, ,
\ee
with constants of integration, $\eta_0$ and $\eta_1$. 
For simplicity, we only consider the case that $\eta_0=\eta_1=0$. 
We also suppose 
$f(\eta)$ is given by
\be
\label{NLdS2}
f(\eta)=f_0 \e^{\frac{\eta}{\beta}}= f_0 \e^{-\frac{4H_0 t}{\beta}}\, , 
\ee
where $f_0$ and $\beta$ are constants. 
In this case, Eq.~(\ref{nl8b}) can be solved as follows, 
\be
\label{NLdS3}
\xi= - \frac{3f_0 \beta}{3\beta - 4} \e^{-\frac{4H_0 t}{\beta}} + \frac{\xi_0}{3H_0}\e^{-3H_0 t} - \xi_1\, , 
\ee
where $\xi_0$ and $\xi_1$ are constants. 
For the de Sitter space, $a$ behaves as $a=a_0\e^{H_0 t}$, 
where $a_0$ is a constant. 
Hence, for the matter with the constant equation of state 
$w_{\mathrm{m}} \equiv P_{\mathrm{m}}/\rho_{\mathrm{m}}$, 
we find
\be
\label{NN1}
\rho_{\mathrm{m}} = \rho_{\mathrm{m} 0} 
\e^{-3(w_{\mathrm{m}}+1)H_0 t}\, , 
\ee
where $\rho_{\mathrm{m} 0}$ is a constant. 
By substituting Eqs.~(\ref{NLdS1}), 
(\ref{NLdS3}) with putting $\xi_0=0$, 
and (\ref{NN1}) into Eq.~(\ref{nl7a}), we obtain 
\be
\label{NLdS4}
0 = - 3H_0^2 \left(1 + \xi_1\right) + 6H_0^2 f_0 \left( \frac{2}{\beta} 
 - 1 \right)\e^{-\frac{4H_0 t}{\beta}} + \Lambda
 + \kappa^2  \rho_{\mathrm{m} 0} \e^{-3(w_{\mathrm{m}}+1)H_0 t} \, . 
\ee
For $\rho_{\mathrm{m} 0}=0$, if we choose
\be
\label{NLdS5}
\beta = 2\ ,\quad 
\xi_1 = - 1 + \frac{\Lambda}{3H_0^2} \, ,
\ee
de Sitter space can be a solution. 
Even if $\rho_{\mathrm{m}} \neq 0$, 
by taking 
\be
\label{NN2}
\beta =\frac{4}{3(1+w_{\mathrm{m}})} \ ,\quad f_0 
= -\frac{\kappa^2 \rho_{\mathrm{m} 0}}{3H_0^2 
\left(1 + 3w_{\mathrm{m}} \right)}\, , 
\quad \xi_1 = -1 + \frac{\Lambda}{3H_0^2} \, ,
\ee 
we find that there is a de Sitter solution. 

The second equation in (\ref{NLdS5}) or third one in (\ref{NN2}) 
implies that 
\be
\label{NNNN1}
H_0^2 = \frac{\Lambda}{3\left( 1 + \xi_1 \right)}\, .
\ee
This means that the cosmological constant $\Lambda$ is effectively 
screened by $\xi$. 
This is one of main results in this paper. 
We include the cosmological term in the action (\ref{nl1}) 
in order to show that the cosmological constant is surely screened by the 
non-local effect. 
We should also note that 
in the case that there does not exist 
the cosmological constant, i.e., $\Lambda=0$, 
if we choose $\xi_1 = -1$, $H_0$ can be arbitrary. 
Thus, $H_0$ can be determined by an initial condition. 
Since $H_0$ can be small or large, the theory with the function in 
Eq.~(\ref{NLdS2}) with $\beta =2$ in the first equation in (\ref{NLdS5}) 
could describe the early-time inflation or current cosmic acceleration. 

In the presence of matter with $w_{\mathrm{m}} \neq 0$, 
for $\Lambda=0$, 
we may have a de Sitter solution $H=H_0$ even if $f(\eta)$ is 
given by 
\be
\label{nLL1}
f(\eta)=f_0\e^{\eta/2} + f_1\e^{3(w_{\mathrm{m}}+1)\eta/4}\ .
\ee
Therefore, the following solution exists:
\be
\label{nLL2}
\eta= - 4H_0 t\ ,\quad \xi=1 - 3f_0\e^{-2H_0t}+ \frac{f_1}{w_{\mathrm{m}}}
\e^{-3(w_{\mathrm{m}}+1)H_0 t}\, , \quad 
\rho_{\mathrm{m}} = - \frac{3(3w_{\mathrm{m}}+1)H_0^2 f_1}{\kappa^2}
\e^{-3(1+w_{\mathrm{m}})H_0t}\, .
\ee

Finally we consider the meaning of the constants of the integration 
$\eta_0$ and $\eta_1$ in (\ref{NLdS1}) and $\xi_0$ and $\xi_1$ in (\ref{NLdS3}). 
First we should note that $\eta_1$ can be absorbed into the shift of the time coordinate 
$t$ as $t\to t + \eta_1/4H_0$ then we can always choose $\eta_0=0$. 
Second as clear from (\ref{NNNN1}), the constant $\xi_1$ determine the initial value 
of scalar curvature $R=12 H_0^2$, in the de Sitter space, where the scalar curvature becomes 
a constant. 
We have chosen $\eta_0=\xi_0=0$ for the technical reason to obtain an exact solution. 
These conditions $\eta_0=\xi_0=0$ give, however, the initial conditions $\eta a^3$, $\xi a^3 \to 0$ 
when $t\to -\infty$. 

\subsection{Condition to be free of ghost}

The existence of a ghost is usually regarded as fatal to
the viability of a theory. But it is not necessarily harmful at the
classical level~\cite{Deruelle:2010kf}. Even at the quantum level, 
it may depend on how one interprets the degree of freedom associated
with a ghost. For example, it has been argued that ghosts may 
be eliminated in the Euclidean path integral approach~\cite{withghost}. 
The ghost surely often generates serious problems in the quantum theory like negative norm etc. 
Even in the classical level, the ghost often generates some problem when we consider the 
cosmological perturbation etc. since the energy of the ghost is unbounded below. 
Therefore the absence of a ghost is certainly a desirable property
of a theory even if we treat the theory classically. 
Here we derive the condition to avoid the appearance of a ghost.

To examine the ghost-free condition,
we make a conformal transformation to the Einstein frame:
\be
\label{E1}
g_{\mu\nu} = \Omega^2 g^{(\mathrm{E})}_{\mu\nu}\, , \quad 
R = \frac{1}{\Omega^2}\left[ R^{(\mathrm{E})} - 6 \left( \Box \ln \Omega 
+ g^{(\mathrm{E})\, \mu\nu} \nabla_\mu \ln \Omega \nabla_\nu \ln \Omega 
\right) \right]\, ,
\ee
with 
\be
\label{E2}
\Omega^2 = \frac{1}{1 + f(\eta) - \xi}\, .
\ee
We use a superscription $(\mathrm{E})$ to represent quantities in 
the Einstein frame. 
Then, the action in Eq.~(\ref{nl2}) gives 
\bea
\label{E3}
S \Eqn{=} 
\int d^4 x \sqrt{ - g^{(\mathrm{E})}} \left\{ \frac{1}{2\kappa^2} 
\left[ R^{(\mathrm{E})} - 6 \left( \Box \ln \Omega 
+ g^{(\mathrm{E})\, \mu\nu} \nabla_\mu \ln \Omega \nabla_\nu \ln \Omega 
\right) \right. \right. \nn
&& \left. \left. 
 - \Omega^2 g^{\mu\nu} \nabla_\mu \xi \nabla_\nu \eta - 2\Omega^4 \Lambda 
\right] + \Omega^4 \mathcal{L}_\mathrm{matter} 
\left(Q; \Omega^2 g^{(\mathrm{E})}\right) \right\} \, .
\eea 
The $\Box \ln \Omega$ term may be discarded because it is a total divergence. 
Hence, we obtain
\bea
\label{E4}
S \Eqn{=}
\int d^4 x \sqrt{ - g^{(\mathrm{E})}} \left[ \frac{1}{2\kappa^2} 
\left( R^{(\mathrm{E})} - 6 g^{(\mathrm{E})\, \mu\nu} \nabla_\mu \ln \Omega 
\nabla_\nu \ln \Omega  
 - \Omega^2 g^{\mu\nu} \nabla_\mu \xi \nabla_\nu \eta - 2\Omega^4 \Lambda 
\right) \right. \nn
&& \left. + \Omega^4 \mathcal{L}_\mathrm{matter} 
\left(Q; \Omega^2 g^{(\mathrm{E})}\right) 
\right] \nn 
\Eqn{=}
\int d^4 x \sqrt{ - g^{(\mathrm{E})}} \left[ \frac{1}{2\kappa^2} \left( 
R^{(\mathrm{E})} - 6 g^{(\mathrm{E})\, \mu\nu} \nabla_\mu \phi \nabla_\nu \phi 
 - \e^{2\phi} g^{\mu\nu} \nabla_\mu \xi \nabla_\nu \eta - 2\e^{4\phi} \Lambda 
\right) \right. \nn
&& \left. + \e^{4\phi} \mathcal{L}_\mathrm{matter} 
\left(Q; \e^{2\phi} g^{(\mathrm{E})}\right) 
\right] \, ,
\eea
where
\be
\label{E5}
\phi = \ln \Omega = - \frac{1}{2} \ln \left( 1 + f(\eta) - \xi \right)\, .
\ee
Instead of $\eta$ and $\xi$, we may regard $\phi$ and $\eta$ to be independent fields. 
Inserting
\be
\label{E6}
\xi = -\e^{-2\phi} + \left( 1 + f(\eta) \right) 
\ee
into the action in Eq.~(\ref{E4}), we finally arrive at
\bea
\label{E7}
S \Eqn{=} 
\int d^4 x \sqrt{ - g^{(\mathrm{E})}} \left\{ \frac{1}{2\kappa^2} \left[ 
R^{(\mathrm{E})} - 6 \nabla^\mu \phi \nabla_\mu \phi - 2 \nabla^\mu \phi 
\nabla_\mu \eta
 - \e^{2\phi} f'(\eta) \nabla^\mu \eta \nabla_\mu \eta
 - 2\e^{4\phi} \Lambda \right] \right. \nn
&& \left. + \e^{4\phi} \mathcal{L}_\mathrm{matter} 
\left(Q; \e^{2\phi} g^{(\mathrm{E})}\right) \right\} \, .
\eea
In order to avoid a ghost, the determinant of the kinetic term must be 
positive, which means~\cite{Nojiri:2010pw} 
\be
\label{E8}
\det \left| \begin{array}{cc} 6 & 1 \\ 1 & \e^{2\phi} f'(\eta) \end{array} \right| 
= 6\e^{2\phi} f'(\eta) -1 > 0\, .
\ee
This condition is assumed to be satisfied. 
In particular, $f'(\eta)>0$ is a 
necessary condition. 
Thus, the ghost-free condition is expressed as 
\begin{equation} 
f'(\eta) > \frac{1}{6\e^{2\phi}} = \frac{1 + f(\eta) - \xi}{6} >0\,, 
\label{eq:2.A-1}
\end{equation}
where in the equality of the middle term, we have used Eq.~(\ref{E5}). 
Then, we may introduce a new field $\chi$, given by 
\be
\label{E9}
\chi = \int^\eta \sqrt{f'(\eta)} d\eta\, ,
\ee
and rewrite the action in Eq.~(\ref{E7}) in the form,
\bea
\label{E10}
S \Eqn{=} 
\int d^4 x \sqrt{ - g^{(\mathrm{E})}} \left\{ \frac{1}{2\kappa^2} \left[ 
R^{(\mathrm{E})} - 6 \nabla^\mu \phi \nabla_\mu \phi - \frac{2}{\sqrt{f'}} 
\nabla^\mu \phi \nabla_\mu \chi
 - \e^{2\phi} \nabla^\mu \chi \nabla_\mu \chi
 - 2\e^{4\phi} \Lambda \right] \right. \nn
&& \left. + \e^{4\phi} \mathcal{L}_\mathrm{matter} 
\left(Q; \e^{2\phi} g^{(\mathrm{E})}\right) \right\} \, ,
\eea
where $f'$ should be regarded as a function 
of $\chi$, $f'(\eta) = f'\left(\eta\left(\chi\right) \right)$. 

In case of the model in Eq.~(\ref{NLdS2}) with $\beta=2$ as in (\ref{NLdS5}), 
for the solution in Eq.~(\ref{NLdS1}) with $\eta_0 = \eta_1=0$ 
and in Eq.~(\ref{NLdS3}) with $\xi_0=0$ and $\xi_1$ given in (\ref{NLdS5}), 
the condition (\ref{E8}) to avoid a ghost has the following form:
\be
\label{ghost1}
\frac{3}{4 + \frac{\Lambda}{3H_0^2 f_0} \e^{2H_0 t}}>1\, ,
\ee
which suggests the two cases. 
One is $f_0$ is positive but the bare cosmological constant $\Lambda$ 
is negative and therefore $\xi_1<-1$ from (\ref{NLdS5}). 
Another is $f_0$ is negative and $\Lambda$ is positive 
and therefore $\xi_1>-1$. 
We note that in the model in Eq.~(\ref{NLdS2}) with $\beta=2$ 
as in (\ref{NLdS5}), 
$f'(\eta) = \left(f_0/2\right) \e^{\eta/2}$ and from 
a necessary condition in order to avoid a ghost $f'(\eta)>0$ 
we find $f_0 >0$, which corresponds to the former case. 
In this case, 
the condition (\ref{ghost1}) suggests that the de Sitter universe is stable 
in a period 
\be
\label{ghost2}
\frac{1}{2H_0}\ln \left( - \frac{3H_0^2 f_0}{\Lambda} \right) < t 
< \frac{1}{2H_0}
\left[\ln4+\ln \left( - \frac{3H_0^2 f_0}{\Lambda} \right)\right] \,, 
%
%
\ee
where we have also used the condition $\left(1 + f(\eta) - \xi\right)/6 >0$ in 
(\ref{eq:2.A-1}). 
Hence the length of the ghost-free period is given by 
\begin{eqnarray}
\Delta t=\frac{\ln4}{2H_0}=\frac{\ln2}{H_0}\simeq\frac{0.69}{H_0}\,.
\end{eqnarray}
Thus unfortunately the period is less than one $e$-folding time. 
So this cannot give inflation in the early universe provided that
the appearance of a ghost has to be avoided.

We note that 
if $f_0$ is negative and $\Lambda$ is positive, 
from the condition (\ref{ghost1}) we see that the de Sitter universe is stable 
when 
$t > \left[ 1/\left(2H_0\right) \right] 
\left[\ln4+\ln \left( - 3H_0^2 f_0/\Lambda \right) \right]$. 
However, this case cannot be available because 
the necessary condition in order to avoid a ghost 
$f'(\eta) = \left(f_0/2\right) \e^{\eta/2} > 0$ is not 
satisfied. 
We also remark that in Ref.~\cite{Elizalde:2011su}, 
it has recently been shown that 
a de Sitter solution exists for an arbitrary value of $\xi_0$. 

\subsection{Cosmology in the Einstein frame}

We explore cosmology in the Einstein frame 
and demonstrate that not only the non-phantom (quintessence) phase 
but also the phantom one can be realized 
in the framework of non-local gravity. 
We assume the FLRW metric (\ref{nl6}), and consider the case
when the contribution of matter is negligible. 

In the absence of matter, the equations given 
by the variation of the action in Eq.~(\ref{E7}) in terms 
of $\phi$ and $\eta$  have the following forms: 
\bea
\label{ms17}
0 \Eqn{=} 
12 \left( \ddot \phi + 3 H \dot \phi \right) + 2 \left( \ddot \eta + 3 H \dot \eta \right) 
 - 2\e^{2\phi} f'(\eta) {\dot \eta}^2 + 8 \e^{4\phi} \Lambda \, , \\
\label{ms18}
0 \Eqn{=} 
2 \left( \ddot \phi + 3 H \dot \phi \right) + 2 \left(\frac{d}{dt} + 3H\right) 
\left( \e^{2\phi} f'(\eta) \dot \eta \right) - \e^{2\phi} f''(\eta) 
{\dot \eta}^2 
\, . 
\eea
We also find that the first FLRW equation is given by 
\be
\label{ms19}
3 H^2 = 3 {\dot\phi}^2 + \dot\phi \dot\eta + \frac{\e^{2\phi}}{2} f'(\eta) 
{\dot\eta}^2  
+ \e^{4\phi} \Lambda \, .
\ee 
We investigate the case that 
\be
\label{MS01}
f'(\eta) = \frac{f_0}{\beta} 
\e^{\frac{\eta}{\beta}}\, .
\ee
We suppose 
\be
\label{MS02}
H = \frac{h_0}{t}\, , \quad 
\phi = - \frac{1}{2} \ln \frac{t}{t_0}\, , \quad 
\eta = \beta \ln \frac{t}{t_0}\, ,
\ee
where $h_0$ and $t_0$ are constants. 
{}From the first equation in (\ref{MS02}), we have $a = a_0 t^{h_0}$, 
where $a_0$ is a constant. 
In this case, Eqs.~(\ref{ms17}), (\ref{ms18}) and (\ref{ms19}) are 
reduced to the algebraic equations
\bea
\label{MS03}
0 \Eqn{=} 
\left( -1 + 3 h_0 \right) \left( 6 - 2 \beta \right) - 2 f_0 \beta
+ 8 t_0^2 \Lambda\, , \\
\label{MS04}  
0 \Eqn{=} 
\left( -1 + 3 h_0 \right) \left( 1 - 2 f_0 \right) - f_0 \, , \\
\label{MS05}
0 \Eqn{=} 
 - 3 h_0^2 + \frac{3}{4} + \frac{\beta}{2} - f_0 \beta \, .
\eea
Then, we can solve $f_0$, $\beta$, and $t_0^2 \Lambda$ as follows
\bea
\label{MS06}
&& f_0 = \frac{1 - 3 h_0}{3\left(1 - 2 h_0 \right)}\, ,\\ 
\label{MS07}
&& \beta = - \frac{ 9 \left( 1 - h_0 \right) \left(3 - 4 h_0 \right) h_0}{2 
\left( 1 - 9 h_0 + 9 h_0^2 \right)}\, ,\\
\label{MS07B}
&& t_0^2 \Lambda = \frac{ 3 \left( -1 + 4 h_0^2 - 9 h_0^3 + 9 h_0^4 \right)}{4 
\left( 1 - 9 h_0 + 9 h_0^2 \right)}
\eea
Then arbitrary $h_0$, except the cases $h_0=1/2$ or $1 - 9 h_0 + 9 h_0^2=0$, 
that is, $h_0 = \frac{ 9 \pm 5\sqrt{3}}{18}$, 
can be realized by choosing $f_0$ and $\beta$ properly if $\Lambda \neq 0$. 
Note that the value of $\Lambda$ can be changed by the redefinition of $t_0$.  

The effective equation of state 
for the universe is given 
by~\cite{Review-Nojiri-Odintsov} 
\begin{equation} 
w_{\mathrm{eff}} \equiv \frac{P_{\mathrm{eff}}}{\rho_{\mathrm{eff}}} = 
-1 - \frac{2\dot{H}}{3H^2} = -1+\frac{2}{3h_0}\,,
\label{eq:2.B-1}
\end{equation}
with 
\begin{eqnarray}
\rho_{\mathrm{eff}} \Eqn{=} \frac{3 H^2}{\kappa^2}\,,
\label{eq:2.C-1} \\ 
P_{\mathrm{eff}} \Eqn{=} -\frac{2\dot H + 3H^2}{\kappa^2}\,.
\label{eq:2.C-2} 
\end{eqnarray}
where 
$\rho_{\mathrm{eff}}$ and $P_{\mathrm{eff}}$ are 
the effective energy density and pressure of the universe, respectively. 
Moreover, 
in the last equality in Eq.~(\ref{eq:2.B-1}), we have used the first equation 
in (\ref{MS02}). 
If $\dot{H} = -h_0/t^2 < 0$, i.e., $h_0 > 0$, we have 
$w_\mathrm{eff} > -1$, which represents the non-phantom (quintessence)
phase, 
while for $\dot{H} = -h_0/t^2 > 0$, i.e., $h_0 < 0$, 
we find 
$w_\mathrm{eff} < -1$, which describes the phantom phase. 
Since almost arbitrary $h_0$ can be realized, there can exist both cases of 
$h_0 > 0$ and $h_0 < 0$. 
We should note that Eq.~(\ref{MS06}) can be rewritten as follows,
\be
\label{h0f0}
h_0 = \frac{1-3f_0}{3(1-2f_0)}\, .
\ee
Therefore if we fix the model $h_0$ is uniquely determined.


We also remark that since we now consider the case in which 
the contribution of matter is absent, 
$w_{\mathrm{eff}}$ in Eq.~(\ref{eq:2.B-1}) is 
equivalent to the equation of state for dark energy. 
This is because 
$\rho_{\mathrm{eff}}$ and $P_{\mathrm{eff}}$ correspond to 
$\rho_{\mathrm{tot}}$ and $P_{\mathrm{tot}}$, respectively, 
where 
$\rho_{\mathrm{tot}}$ and $P_{\mathrm{tot}}$ 
are the total energy density and pressure of the universe, 
respectively, and thus 
$w_{\mathrm{eff}}$ can be expressed as 
$w_{\mathrm{eff}} = P_{\mathrm{tot}}/\rho_{\mathrm{tot}}$. 

Furthermore, 
since cosmological screening is a significant result of the paper, it 
will be useful to explain the mechanism in the Einstein frame. 
It follows from the form of the potential term for 
the scalar field $\phi$ in the action in Eq.~(\ref{E7}) that 
$\phi$ runs towards negative infinity with time. 
Presumably, thereby screening the cosmological constant can be realized. 
It would be good to see this point explicitly. 
For clear understanding, we define the dimensionless quantities as 
$\tilde{\Lambda} \equiv t_0^2 \Lambda$ and $\tilde{H}_0 \equiv H (t = t_0) t_0 
= h_0$, where the second equality follows from the first equation 
in (\ref{MS02}). {}From the solution in Eq.~(\ref{MS07B}), in principle, 
by solving Eq.~(\ref{MS07B}) reversely, we obtain the expression of 
$\tilde{H}_0^2$ as a function of $\tilde{\Lambda}$ as 
\begin{equation}  
\tilde{H}_0^2 = \tilde{H}_0^2 (\tilde{\Lambda})\,. 
\label{eq:Add-SecIIC-01} 
\end{equation}
%
%
In the second equation $\phi = - \left(1/2\right) \ln \left(t/t_0\right)$ in 
(\ref{E7}), $\phi = 0$ at $t=t_0$. In other words, the definition in terms of 
the value of $t_0$ can be determined by the dynamics (or expression) of 
$\phi$. 
Thus, it can be considered that 
Eq.~(\ref{eq:Add-SecIIC-01}) implies that the cosmological constant 
$\Lambda$ is effectively screened by $\phi$, 
similarly to that in Eq.~(\ref{NNNN1}) in the Jordan frame. 
Incidentally, a similar mechanism was proposed in Ref.~\cite{Biswas:2005vz} 
to address the hierarchy problem. 

In addition, 
there are strong experimental constraints on violation of 
equivalence principle and the existence of a fifth force mediated by scalars, 
which is easily seen in the Einstein frame. In the non-local model, 
the scalar field $\phi$ couples to the matter sector and this could be a 
serious problem. 
Essentially, what needs to be very small is 
$\left( d \Omega/ d t \right) /\Omega$. 
It is necessary to check this before presenting the model as a serious 
candidate for dark energy. 
For the solution $\phi = - \left(1/2\right) \ln \left(t/t_0\right)$ in 
the second equation in (\ref{E7}), we find 
$\left( d \Omega/ d t \right) /\Omega = \dot{\phi} = 
-1/\left(2 t \right)$. 
It is clear that in the limit of $t \to \infty$, 
$\dot{\phi} \to 0$. 
For example, this limit can be regarded as the present time. 
Consequently, the rate of the time variation of $\Omega$ can be 
very small at the current universe. 

\subsection{Addition of an $R^2$ term}

We examine the influence of adding an $R^2$ term 
on the stability of non-local gravity 
in the Einstein frame. The addition of it is motivated by
the possible curing of the finite-time future singularities. 
We study the case in which 
$u R^2/\left(2\kappa^2\right)$ with $u (\neq 0)$ being a non-zero 
arbitrary constant is added to the action in Eq.~(\ref{nl1}) as follows: 
\begin{equation}  
S=\int d^4 x \sqrt{-g}\left\{
\frac{1}{2\kappa^2}\left[ R\left(1 + f(\Box^{-1}R )\right) 
+ u R^2 
-2 \Lambda \right]
+ \mathcal{L}_\mathrm{matter} \left(Q; g\right)
\right\}
\,.
\label{eq:2-F-1} 
\end{equation}
We introduce another scalar field $\zeta$. 
The action in Eq.~(\ref{eq:2-F-1}) can be rewritten to 
\begin{equation}
S = 
\int d^4 x \sqrt{-g}\left\{
\frac{1}{2\kappa^2}\left[R\left(1 + f(\eta)\right) 
 - \partial_\mu \xi \partial^\mu \eta - \xi R 
+ u\left( 2 \zeta R - \zeta^2 \right) 
- 2 \Lambda \right]
+ \mathcal{L}_\mathrm{matter} 
\right\}\,.
\label{eq:2-F-2} 
\end{equation}
By varying the action in Eq.~(\ref{eq:2-F-2}) with respect to 
$\zeta$, we have $\zeta=R$. 
Substituting this equation into the action in Eq.~(\ref{eq:2-F-2}), 
the action in Eq.~(\ref{eq:2-F-1}) is re-obtained.

When an $R^2$ term is added, i.e., 
for the action in Eq.~(\ref{eq:2-F-1}), 
in the flat FLRW background in Eq.~(\ref{nl6}), 
gravitational field equations are explicitly written down as 
\begin{eqnarray}
\hspace{-11mm}
0 \Eqn{=} 
- 3 H^2\left(1 + f(\eta) - \xi\right) + \frac{1}{2}\dot\xi \dot\eta 
 - 3H\left(f'(\eta)\dot\eta - \dot\xi\right) 
+ \Theta 
+ \Lambda 
+ \kappa^2 \rho_{\mathrm{m}}\,,
\label{eq:2-F-7} \\ 
\hspace{-11mm}
0 \Eqn{=} 
\left(2\dot H + 3H^2\right) \left(1 + f(\eta) - \xi\right) 
+ \frac{1}{2}\dot\xi \dot\eta 
+ \left(\frac{d^2}{dt^2} + 2H \frac{d}{dt} \right) \left( f(\eta) - 
\xi \right) 
+ \Xi 
- \Lambda + \kappa^2 P_{\mathrm{m}}\,, 
\label{eq:2-F-8} 
\end{eqnarray}
where $\Theta$ and $\Xi$ are the contributions from the additional 
term $u R^2/\left(2\kappa^2\right)$, 
given by 
\begin{eqnarray} 
\Theta \Eqn{\equiv} u \left( -6H^2 R + \frac{1}{2}R^2 -6H\dot{R} \right) 
= 18u\left( -6H^2 \dot{H} + \dot{H}^2 -2H \ddot{H} \right)\,, 
\label{eq:2-F-9} \\
\Xi \Eqn{\equiv} 
u \left[ 
2 \left(2\dot{H} + 3H^2\right) R 
-\frac{1}{2}R^2 +2\ddot{R} +4H\dot{R} 
\right] 
\nonumber \\
\Eqn{=} 
6u\left( 9\dot{H}^2 +18H^2 \dot{H} + 2\dddot{H} + 12H\ddot{H} \right)\,. 
\label{eq:2-F-10} 
\end{eqnarray} 
Namely, $\Theta$ is added to the r.h.s. of Eq.~(\ref{nl7a}) 
and $\Xi$ is added to the r.h.s. of Eq.~(\ref{nl7b}). 
Here, 
in deriving the second equalities in Eqs.~(\ref{eq:2-F-9}) and 
(\ref{eq:2-F-10}) we have used $R = 6\dot{H} + 12H^2$. 
{}From Eqs.~(\ref{eq:2-F-9}) and (\ref{eq:2-F-10}), we see that 
all the terms in $\Theta$ and $\Xi$ include the derivative terms of $H$ with 
respect to time. 
Hence, for a de Sitter solution $H=H_0$, $\Theta$ and $\Xi$ 
vanish and therefore do not change the de Sitter solutions. 
As a result, 
the solutions in (\ref{NN2}) and (\ref{nLL2}) can also be obtained 
in the presence of the additional $R^2$ term. 

By using Eqs.~(\ref{nl7a}), (\ref{nl7b}), (\ref{eq:2.B-1}), (\ref{eq:2.C-1}) 
and (\ref{eq:2.C-2}), 
the effective equation of state for the universe is given by 
\begin{equation} 
w_{\mathrm{eff}} = \frac{P_{\mathrm{eff}}}{\rho_{\mathrm{eff}}} 
= \frac{\left(2\dot H + 3H^2\right) \left( f(\eta) - \xi \right) 
+ \frac{1}{2}\dot\xi \dot\eta 
+ \left(\frac{d^2}{dt^2} + 2H \frac{d}{dt} \right) \left( f(\eta) - 
\xi \right) - \Lambda + \kappa^2 P_{\mathrm{m}}}{
- 3 H^2\left( f(\eta) - \xi \right) 
+ \frac{1}{2}\dot\xi \dot\eta 
 - 3H\left(f'(\eta)\dot\eta - \dot\xi\right) + \Lambda 
+ \kappa^2 \rho_{\mathrm{m}}}\,,
\label{eq:5.1}
\end{equation}
where
\begin{eqnarray}
\hspace{-10mm} 
\rho_{\mathrm{eff}} \Eqn{=} 
\frac{1}{\kappa^2} 
\left[ 
- 3 H^2\left( f(\eta) - \xi \right)
+ \frac{1}{2}\dot\xi \dot\eta 
 - 3H\left(f'(\eta)\dot\eta - \dot\xi\right) + \Lambda 
+ \kappa^2 \rho_{\mathrm{m}}
\right]\,,
\label{eq:5.2} \\ 
\hspace{-10mm} 
P_{\mathrm{eff}} \Eqn{=} 
\frac{1}{\kappa^2} 
\left[ \left(2\dot H + 3H^2\right) \left( f(\eta) - \xi \right) 
+ \frac{1}{2}\dot\xi \dot\eta 
+ \left(\frac{d^2}{dt^2} + 2H \frac{d}{dt} \right) \left( f(\eta) - 
\xi \right) - \Lambda + \kappa^2 P_{\mathrm{m}}
\right]\,.
\label{eq:5.3} 
\end{eqnarray}
If we add an $R^2$ term as in the action in Eq.~(\ref{eq:2-F-1}), 
$\rho_{\mathrm{eff}}$ and $P_{\mathrm{eff}}$ become 
\begin{eqnarray}
\hspace{-10mm} 
\rho_{\mathrm{eff}} \Eqn{=} 
\frac{1}{\kappa^2} 
\left[ 
- 3 H^2\left( f(\eta) - \xi \right)
+ \frac{1}{2}\dot\xi \dot\eta 
 - 3H\left(f'(\eta)\dot\eta - \dot\xi\right) + \Theta + \Lambda 
+ \kappa^2 \rho_{\mathrm{m}}
\right]\,,
\label{eq:Addition-2-E-001} \\ 
\hspace{-10mm} 
P_{\mathrm{eff}} \Eqn{=} 
\frac{1}{\kappa^2} 
\left[ \left(2\dot H + 3H^2\right) \left( f(\eta) - \xi \right) 
+ \frac{1}{2}\dot\xi \dot\eta 
+ \left(\frac{d^2}{dt^2} + 2H \frac{d}{dt} \right) \left( f(\eta) - 
\xi \right) \right. \nonumber \\ 
&& \left. 
{}+ \Xi - \Lambda + \kappa^2 P_{\mathrm{m}}
\right]\,.
\label{eq:Addition-2-E-001BB} 
\end{eqnarray}
%
For a de Sitter solution $H=H_0$, 
since $\Theta$ and $\Xi$ vanish, 
$w_{\mathrm{eff}}$ does not change in the presence of an $R^2$ term.
In Sec.~V C, we discuss the cosmological consequences of 
adding an $R^2$ term for more general solutions. 
In particular, it is shown that an $R^2$ term plays a fundamental
role to cure the finite-time future singularities in non-local gravity.

Let us examine the condition to avoid a ghost in the present case. 
By following the same procedure in Sec.~II C, 
we perform a conformal transformation to the Einstein frame 
as in (\ref{E1}) with 
\begin{equation}
\Omega^2 = \frac{1}{1 + f(\eta) - \xi + 2 u \zeta}\,. 
\label{eq:2-F-3} 
\end{equation}
In this case, we find 
\begin{eqnarray}
S 
\Eqn{=}
\int d^4 x \sqrt{ - g^{(\mathrm{E})}} \left[ \frac{1}{2\kappa^2} \left( 
R^{(\mathrm{E})} - 6 g^{(\mathrm{E})\, \mu\nu} \nabla_\mu \phi \nabla_\nu \phi 
 - \e^{2\phi} g^{\mu\nu} \nabla_\mu \xi \nabla_\nu \eta 
 - u \e^{4\phi} \zeta^2 
 - 2\e^{4\phi} \Lambda 
\right) \right. \nn
&& \left. + \e^{4\phi} \mathcal{L}_\mathrm{matter} 
\left(Q; \e^{2\phi} g^{(\mathrm{E})}\right) 
\right]\,,
\label{eq:2-F-4} 
\end{eqnarray}
where
$
\phi = \ln \Omega = - \left( 1/2 \right) 
\ln \left( 1 + f(\eta) - \xi + 2 u \zeta \right). 
$
Substituting 
$
\xi = -\e^{-2\phi} + \left( 1 + f(\eta) \right) + 2 u \zeta 
$ 
into the action in Eq.~(\ref{eq:2-F-4}), we acquire 
\begin{eqnarray}
S \Eqn{=} 
\int d^4 x \sqrt{ - g^{(\mathrm{E})}} \left[ \frac{1}{2\kappa^2} \left( 
R^{(\mathrm{E})} - 6 \nabla^\mu \phi \nabla_\mu \phi - 2 \nabla^\mu \phi 
\nabla_\mu \eta
- \e^{2\phi} f'(\eta) \nabla^\mu \eta \nabla_\mu \eta 
\right. 
\right. \nn
&& \left. 
\left. 
- 2u \e^{2\phi} \nabla^\mu \zeta \nabla_\mu \eta 
- u \e^{4\phi} \zeta^2 
- 2\e^{4\phi} \Lambda \right) 
+ \e^{4\phi} \mathcal{L}_\mathrm{matter} 
\left(Q; \e^{2\phi} g^{(\mathrm{E})}\right) \right] \,.
\label{eq:2-F-5} 
\end{eqnarray}
In this case, the matrix for the kinetic terms is given by 
\begin{equation} 
M \equiv 
\left( 
\begin{array}{ccc} 
6 & 1 & u \e^{2\phi} \\ 
1 & \e^{2\phi} f'(\eta) & 0 \\
u \e^{2\phi} & 0 & 0 \\
\end{array} 
\right)\,.
\label{eq:ADD-2F-1} 
\end{equation}
The necessary condition to avoid a ghost is that 
all the eigenvalues of the matrix for the kinetic terms $M$ must be positive. 
The characteristic equation for $M$ is given by 
\begin{equation} 
\det \left| 
M - y E 
\right| 
= 
\det \left| 
\begin{array}{ccc} 
6 - y & 1 & u \e^{2\phi} \\ 
1 & \e^{2\phi} f'(\eta) - y & 0 \\
u \e^{2\phi} & 0 & - y \\
\end{array} 
\right| 
= 0\,, 
\label{eq:ADD-2F-2} 
\end{equation}
where $y$ denotes an eigenvalue of $M$ and $E$ is a unit matrix. 
From Eq.~(\ref{eq:ADD-2F-2}), we obtain 
\begin{equation} 
y^3 - 
Y_1 y^2 
+ 
Y_2 y 
- 
Y_3 = 0\,, 
\label{eq:ADD-2F-3} 
\end{equation}
with 
\begin{eqnarray} 
Y_1 \Eqn{\equiv} 
y_1 + y_2 + y_3 
= 
\e^{2\phi} f'(\eta) + 6\,, 
\label{eq:ADD-2F-4} \\
Y_2 \Eqn{\equiv} 
y_2 y_3 + y_3 y_1 + y_1 y_2 
= 6 \e^{2\phi} f'(\eta) - u^2 \e^{4\phi} - 1\,, 
\label{eq:ADD-2F-5} \\
Y_3 \Eqn{\equiv} 
y_1 y_2 y_3 
= - u^2 \e^{6\phi} f'(\eta)\,, 
\label{eq:ADD-2F-6} 
\end{eqnarray} 
where $y_1$, $y_2$ and $y_3$ are three eigenvalues of $M$. 
In order to avoid a ghost, we must have 
$y_1 > 0$, $y_2 > 0$ and $y_3 > 0$, 
and therefore 
$Y_1 > 0$, $Y_2 > 0$ and $Y_3 > 0$. 
If $f'(\eta) < 0$, $Y_3 > 0$. 
Moreover, if $- 6 \e^{-2\phi} < f'(\eta)$, $Y_1 > 0$. 
However, when $f'(\eta) < 0$, $Y_2 < 0$. This implies that 
all the three eigenvalues of $M$ cannot be positive simultaneously. 
Thus, unfortunately the necessary condition to avoid a ghost cannot be 
satisfied. It may be noted, however, that the addition of a non-minimally 
coupled scalar instead of an $R^2$ term may help 
in the realization of a solution without a
ghost~\cite{Zhang:2011uv}. 
We note that the conditions $Y_1 > 0$ and $Y_3 > 0$ correspond to 
$\mathrm{Tr} \, M >0$ and $\det \left| M \right| >0$, respectively. 

If we add the general $F(R)$ term instead of a simple $R^2$ term and, 
as usual, by rewriting the action as~\cite{Review-Nojiri-Odintsov} 
$F(R)$ to $F(A) + F'(A) (R - A)$, where 
$A$ is an auxiliary field and $F'(A) \equiv dF(A)/dA$, 
we may redo the analysis parallel to the present $R^2$ case. 
Clearly, however, if we introduce a scalar field $\zeta$ by 
setting $\zeta=F'(A)/(2u)$, the ghost-free condition does not change at all.
Hence the addition of an $F(R)$ term does not remove the ghost.
Also note that the previous de Sitter solution would no longer be a solution.

We note that 
the creation of the Standard Model particles is very important 
and serious problem. 
Since the non-local gravity is very complicated model, however, it is very 
important to show the existence of the consistent de Sitter solution, which 
could be applied to the cosmic accelerated expansion in the early universe, 
i.e., inflation. 
In order to generate radiation and non-relativistic matters including 
Standard Model quarks and leptons, it is necessary to consider a kind 
of the existence of the inflaton field, which decays into radiation, 
as a standard inflation scenario, 
and a scenario for the baryogenesis and/or leptogenesis as well as 
a generation mechanism of dark matter. 
These problems of the particle creations consistent with the observation of 
the universe could be investigated in the future works. 

We also mention that 
gravity being universal will couple to visible and hidden sectors, and 
in action Eq.~(\ref{eq:2-F-2}) the couplings will be universal. 
Therefore, it is important to consider how we could explain that after 
inflation solely the Standard Model degrees are excited from the vacuum. 
In addition, Big Bang Nucleosynthesis (BBN) puts stringent constraint on 
any hidden or dark radiation. 
Our discussion in Sec.~II E is an attempt to 
examine the influence of adding an $R^2$ term 
on the stability of non-local gravity in the Einstein frame, 
i.e., the ghost-free condition. 
Thus, it is necessary not only to explore a mechanism to create 
the Standard Model degrees after inflation but also to 
investigate whether the resultant inflationary scenario in non-local gravity 
could satisfy the BBN constraints. 
We like to investigate these problems in future. 

\section{Non-local gravity with Lagrange constraint multiplier}

Recently, in Refs.~\cite{Nojiri:2010tv, Nojiri:2009th} the model of covariant 
gravity with properties similar to Ho\v{r}ava gravity~\cite{Horava:2009uw} was 
introduced, adding Lagrange constraint 
multiplier~\cite{Lagrange-Constraint-Multiplier}. 
In this section, we study the Ho\v{r}ava like extension covariantly proposed 
in Ref.~\cite{Nojiri:2010tv}. 
We generalize non-local gravity by introducing Lagrange constraint 
multiplier and examine a de Sitter solution in non-local gravity with 
Lagrange constraint multiplier~\cite{Nojiri:2010pw}. 

In this section, however, we do not discuss the condition for
the absence of a ghost. This is partly because the analysis would
become much more complicated. But the main reason is that 
in the present case the local Lorentz invariance will be explicitly 
violated in all solutions of the theory. This implies that
the existence of a ghost, which would be fatal for a quatum theory
with Lorentz invariance,  may not be a real problem at all.

The constrained action for a scalar field $\psi$ is given by 
\begin{equation} 
S_{\psi} = \int d^4 x \sqrt{-g} 
\left[ 
- \lambda \left( \frac{1}{2} \partial_\mu \psi \partial^\mu \psi
+ U( \psi) \right) 
\right]\,,
\label{eq:3.A-1}
\end{equation}
where $\lambda$ is the Lagrange multiplier field. 
This action gives a constraint 
$\left(1/2\right) \partial_\mu \psi \partial^\mu \psi
+ U( \psi ) = 0$. 
The vector $\partial_\mu \psi$ is time-like. 
Therefore, the direction of time can be chosen to be parallel to 
$\partial_\mu \psi$ at least locally. Hence, it follows from this 
constraint that $\left(1/2\right) \left(d \psi/dt\right)^2 = U( \psi)$. 
For simplicity, we choose $U( \psi )$ to be a constant as 
$U( \psi ) = U_0$. In this case, we have
\be
\label{LagHL2}
\frac{1}{2} \partial_\mu \psi \partial^\mu \psi
+ U_0 = 0\, . 
\ee
Under the constraint (\ref{LagHL2}), we now define
\bea
\label{nlHrv1}
R^{(2n+2)} 
\Eqn{\equiv} 
R - 2\kappa^2\alpha \left[
\left(\partial^\mu \psi \partial^\nu \psi \nabla_\mu \nabla_\nu
+ 2 U_0 \nabla^\rho \nabla_\rho \right)^n
\left( \partial^\mu \psi \partial^\nu \psi R_{\mu\nu} + U_0 R 
\right)\right]^2 \, , \nn
R^{(2n+3)} 
\Eqn{\equiv} 
R - 2\kappa^2 \alpha \left[
\left(\partial^\mu \psi \partial^\nu \psi \nabla_\mu \nabla_\nu
+ 2 U_0 \nabla^\rho \nabla_\rho \right)^n
\left( \partial^\mu \psi \partial^\nu \psi R_{\mu\nu} + U_0 R \right)\right]  
\nn 
&& \times \left[
\left(\partial^\mu \psi \partial^\nu \psi \nabla_\mu \nabla_\nu
+ 2 U_0 \nabla^\rho \nabla_\rho \right)^{n+1}
\left( \partial^\mu \psi \partial^\nu \psi R_{\mu\nu} + U_0 R 
\right)\right]\, , \nn
\Box^{(n)} 
\Eqn{\equiv} 
\Box + 
\gamma 
\left(\partial^\mu \psi \partial^\nu \psi \nabla_\mu \nabla_\nu
+ 2 U_0 \nabla^\rho \nabla_\rho \right)^n \, , 
\eea
where $n$, $\alpha$ and $\gamma$ are constants. 
In a parallel way from Eq.~(\ref{nl1}), we may define the non-local 
action~\cite{Nojiri:2010pw} 
\bea
\label{nlHrv2}
S \Eqn{=} \int d^4 x \sqrt{-g}\left\{
\frac{1}{2\kappa^2}\left[ R^{(m)}\left(1 + f\left(\left(\Box^{n}\right)^{-1}R^{(m)}\right)\right) 
 - 2\Lambda \right] \right. \nn
&& \left. - \lambda \left( \frac{1}{2} \partial_\mu \psi \partial^\mu \psi
+ U_0 \right) + \mathcal{L}_\mathrm{matter} \right\}\, , 
\eea
and rewrite the action in Eq.~(\ref{nlHrv2}) in a local way by introducing two 
scalar fields $\eta$ 
and $\xi$~\cite{Nojiri:2010pw}:
\bea
\label{nlHrv3}
S \Eqn{=} 
\int d^4 x \sqrt{-g}\left\{
\frac{1}{2\kappa^2}\left[R^{(m)} \left(1 + f(\eta)\right) 
+ \xi\left(\Box^{(n)} \eta - R^{(m)}\right) -2 \Lambda \right]
 - \lambda \left( \frac{1}{2} \partial_\mu \psi \partial^\mu \psi
+ U_0 \right) \right. \nn
&& \left. 
+ \mathcal{L}_\mathrm{matter}  \right\} \, .
\eea
In Eqs.~(\ref{nlHrv2}) and (\ref{nlHrv3}), $n$ can be even or odd integer. 

In order to explore the cosmology, we assume the following 
form of the metric:
\be
\label{bFRW}
ds^2 = - \e^{2b(t)}dt^2 + a^2(t) \sum_{i=1,2,3} \left(dx^i\right)^2\, ,
\ee
and that the scalar field $\psi$ only depends on time. 
Then, we have 
\bea
\label{nlHrv4}
&&
\partial^\mu \psi \partial^\nu \psi R_{\mu\nu} + U_0 R = 6 U_0 \e^{-2b} H^2 \, ,\quad 
\partial^\mu \psi \partial^\nu \psi \nabla_\mu \nabla_\nu
+ 2 U_0 \nabla^\rho \nabla_\rho = - 6 U_0 \e^{-2b} H \partial_t\, , 
\nonumber \\ 
&&
R = \e^{-2b} \left( 6\dot H + 12 H^2 -6\dot b H \right)\, . 
\eea
We suppose the scalar fields $\eta$ and $\xi$ only depend on $t$. 
We examine most simple but non-trivial case that $m=2$, i.e., $n=0$, 
in Eq.~(\ref{nlHrv3}) and $\gamma=0$ in (\ref{nlHrv1}). 
For $m=2$, it follows from the first equation in (\ref{nlHrv1}) with $n=0$ 
that 
\begin{equation}
R^{(2)} 
= R - 2\kappa^2\alpha 
\left( \partial^\mu \psi \partial^\nu \psi R_{\mu\nu} + U_0 R 
\right)^2\,.  
\label{eq:3.A-2}
\end{equation}
%
The variation of the action in Eq.~(\ref{nlHrv3}) 
with respect to $\eta$ gives 
$0=\Box\xi+ f'(\eta) R^{(2)}$. 
In the background in Eq.~(\ref{bFRW}), this equation yields 
\be
\label{nlHrv5}
0 = \left(6\dot H + 12 H^2 - 72\kappa^2 \alpha U_0^2 H^4 \right) f'\left(\eta\right) 
 - \ddot \xi - 3H \dot\xi\, ,
\ee
after putting $b=0$. 
Moreover, the variation of the action in Eq.~(\ref{nlHrv3}) 
with respect to $\xi$ leads to $\Box\eta=R^{(2)}$. 
In the background in Eq.~(\ref{bFRW}), this equation becomes 
\be
\label{nlHrv6}
0 = 6\dot H + 12 H^2 - 72 \kappa^2 \alpha U_0^2 H^4 + \ddot \eta 
+ 3H \dot\eta\, .
\ee
In deriving Eqs.~(\ref{nlHrv5}) and (\ref{nlHrv6}), 
we have used Eq.~(\ref{eq:3.A-2}) and the first and third equations in 
(\ref{nlHrv4}). 

For $m=2$, the action in Eq.~(\ref{nlHrv3}) is expressed as 
\begin{eqnarray} 
S \Eqn{=} \int d^4 x a^3 
\left(
\frac{1}{2\kappa^2}\biggl\{ 
\left[ 
\e^{-b} 
\left( 6\dot{H} + 12 H^2 -6\dot{b} H \right)
 - 72 \kappa^2 \alpha U_0^2 \e^{-3b} H^4 
\right] 
\left(1 + f(\eta) - \xi \right) 
\right. \nonumber \\ 
&& \left. 
+\e^{-b} \dot{\xi} \dot{\eta} -2\e^{b} \Lambda 
\biggr\} 
 - \lambda \left( -\frac{\e^{-b}}{2} \dot{\psi}^2 + \e^{b} U_0 \right) 
+ \mathcal{L}_\mathrm{matter} \right)\,,
\label{eq:3.A-3}
\end{eqnarray}
where we have used the metric in Eq.~(\ref{bFRW}), 
the first and third equations in (\ref{nlHrv4}) 
and Eq.~(\ref{eq:3.A-2}). 

By the variation of the action in Eq.~(\ref{eq:3.A-3}) 
with respect to $b$, 
we obtain the equation corresponding to the first 
FLRW equation: 
\be
\label{nlHrv7}
0 = -  3\left( H^2 + 36 \kappa^2 \alpha U_0^2 H^4 \right) 
\left(1 + f(\eta) - \xi \right) + \frac{1}{2}\dot\xi \dot\eta 
 - 3 H \left(f'(\eta) \dot\eta - \dot\xi \right) + \Lambda 
+ 2 \kappa^2 \lambda U_0 
+ \kappa^2 \rho_\mathrm{m} 
\, ,
\ee
where we have used $\left(1/2\right) \left(d \psi/dt\right)^2 = U_0$. 
Moreover, the variation of the action in Eq.~(\ref{eq:3.A-3}) 
with respect to $a$ gives the equation corresponding to the second 
FLRW equation:
\bea
\label{nlHrv8}
\hspace{-10mm} 
&& 0 =  \left( 2\dot H + 3 H^2 + 108 \kappa^2 \alpha U_0^2 H^4 
+ 144 \kappa^2 \alpha U_0^2 H^2 \dot H \right) 
\left(1 + f ( \eta) - \xi \right) 
+ \frac{1}{2}\dot\xi \dot\eta \nn
\hspace{-10mm} 
&& \qquad + \left( 2H 
+ 48 \kappa^2 \alpha U_0^2 H^3 \right) 
\left(f'(\eta) \dot\eta - \dot\xi \right) 
+ f'' ( \eta) \left(\dot\eta\right)^2 
+ f'(\eta) \ddot\eta - \ddot\xi - \Lambda + \kappa^2 P_\mathrm{m}\, .
\eea
We investigate if there could be a solution describing the de Sitter space.  
When we assume $H$ is a constant, $H=H_0$, Eq.~(\ref{nlHrv6}) could be 
integrated, as executed in (\ref{NLdS1}), 
\be
\label{nlHrv9}
\eta= - 4H_0 \left( 1 - 6 \kappa^2 \alpha U_0^2 H_0^2 \right) t - \eta_0 \e^{-3H_0 t} 
+ \eta_1\, .
\ee
Again we take $\eta_0 = \eta_1 = 0$. 
We suppose $f(\eta)$ is given by (\ref{NLdS2}) 
as $f(\eta)=f_0 \e^{\frac{\eta}{\beta}}$. 
By integrating Eq.~(\ref{nlHrv5}), we obtain
\be
\label{nlHrv10}
\xi = - \frac{3f_0 H_0 t_{\mathrm{c}}}{1 + 3 H_0 t_{\mathrm{c}}} 
\e^{\frac{t}{t_{\mathrm{c}}}} 
+ \frac{\xi_0}{3H_0} \e^{-3H_0 t} - \xi_1\, . 
\ee
Here, $\xi_0$ and $\xi_1$ are constants of the integration. 
Moreover, $t_{\mathrm{c}}$ is given by
\be
\label{nlHrv11}
t_{\mathrm{c}} \equiv 
- \frac{\beta}{4H_0 \left( 1 - 6 \kappa^2 \alpha U_0^2 H_0^2 \right)}\, .
\ee
We now neglect the contribution from matter. 
By substituting Eqs.~(\ref{nlHrv9}) with 
$\eta_0 = \eta_1 = 0$ and (\ref{nlHrv10}) into 
Eq.~(\ref{nlHrv8}), we find
\be
\label{nlHrv12}
\xi_0 = 0\, ,\quad 
\xi_1 = -1 + \frac{\Lambda}{3\left( H_0^2 + 36 \kappa^2 \alpha U_0^2 H_0^4 \right)} 
\, ,
\ee
and 
\be
\label{nllHrv13} 
0 = 18 \left(1 + 36 \kappa^2 \alpha U_0^2 H_0^2 \right) H_0^3 t_{\mathrm{c}}^3 
+ 3 \left(7 + 120 \kappa^2 \alpha U_0^2 H_0^2 \right) H_0^2 t_{\mathrm{c}}^2 
+ 8 \left(1 + 6 \kappa^2 \alpha U_0^2 H_0^2 \right) H_0 t_{\mathrm{c}} + 1\, ,
\ee
which is the cubic algebraic equation with respect to $t_{\mathrm{c}}$. 
Hence, there is always a real solution of $t_{\mathrm{c}}$. 
Thus, we can find the value of $\beta$ by using Eq.~(\ref{nlHrv11}). 
Consequently, 
if we choose $\beta$ properly, there always appears the solution 
describing the de Sitter universe. 

The second equation in (\ref{nlHrv12}) can be rewritten as
\be
\label{NNNNN}
H_0^2 = \frac{\Lambda}{3\left(1 + \xi_1\right)} 
- 36 \kappa^2 \alpha U_0^2 H_0^4 \, .
\ee
Therefore, 
$\xi_1$ can be a screening of the cosmological constant and the last term is a 
correction coming from the covariant Ho\v{r}ava like model. 

As discussed in Sec.~II E, 
the non-local action often appears as a quantum correction. 
Equation (\ref{NNNNN}) implies 
again that the correction effectively changes the 
value of the cosmological constant. 
If $\xi\sim 0$ in the early universe, where $t\sim 0$, Eq.~(\ref{nlHrv10}) 
means 
\be
\label{NNNNN2} 
\xi_1 \sim - \frac{3f_0 H_0 t_{\mathrm{c}}}{1 + 3 H_0 t_{\mathrm{c}}} \, .
\ee
If $H_0$ corresponds to the value of the Hubble parameter 
in the present universe, 
the second term on the right-hand side (r.h.s.) 
of Eq.~(\ref{NNNNN}) could be negligible. 
Thus, if $- \frac{3f_0 H_0 t_{\mathrm{c}}}{1 + 3 H_0 t_{\mathrm{c}}}$ is 
positive and very large, the effective 
cosmological constant in the present universe could be very small.

We explore a condition to avoid the appearance of a ghost by 
following the procedure in Sec.~II C. 
We make a conformal transformation 
in (\ref{E1}) to the Einstein frame: 
\begin{equation} 
g_{\mu\nu} = \Omega^2 g^{(\mathrm{E})}_{\mu\nu}\, , \quad 
R^{(2)} = \frac{1}{\Omega^2}\left[ R^{(2,\, \mathrm{E})} 
- 6 \left( \Box \ln \Omega 
+ g^{(\mathrm{E})\, \mu\nu} \nabla_\mu \ln \Omega \nabla_\nu \ln \Omega 
\right) \right]\,, 
\label{eq:3.B-1} 
\end{equation}
with Eq.~(\ref{E2}). 
We here describe the Lagrangian for the part of the Lagrange multiplier field 
as 
\begin{equation} 
\mathcal{L}_{\lambda} 
\left(\Upsilon; g\right) 
= 
- \lambda \left( \frac{1}{2} \partial_\mu \psi \partial^\mu \psi
+ U_0 \right)\,, 
\label{eq:3.B-2} 
\end{equation}
where $\Upsilon$ denotes the Lagrange multiplier field $\lambda$ and 
the scalar field $\psi$. 
Through the conformal transformation in (\ref{eq:3.B-1}), 
the action in the Einstein frame is expressed as 
\begin{eqnarray}
S 
\Eqn{=} 
\int d^4 x \sqrt{ - g^{(\mathrm{E})}} \left[ \frac{1}{2\kappa^2} \left( 
R^{(2,\, \mathrm{E})} - 6 \nabla^\mu \phi \nabla_\mu \phi - 2 \nabla^\mu \phi 
\nabla_\mu \eta
 - \e^{2\phi} f'(\eta) \nabla^\mu \eta \nabla_\mu \eta
 - 2\e^{4\phi} \Lambda \right) \right. \nn
&& \left. 
+ \e^{4\phi} \mathcal{L}_{\lambda} 
\left(\Upsilon; \e^{2\phi} g^{(\mathrm{E})}\right) 
+ \e^{4\phi} \mathcal{L}_\mathrm{matter} 
\left(Q; \e^{2\phi} g^{(\mathrm{E})}\right) \right] 
\,. 
\label{eq:3.B-3}
\end{eqnarray}
Since the form of the action in Eq.~(\ref{eq:3.B-3}) is 
equivalent to that in Eq.~(\ref{E7}) 
except the Lagrangian for the part of the Lagrange multiplier field 
$\mathcal{L}_{\lambda}$, 
a condition to avoid a ghost is given by (\ref{E8}). 
We also have a necessary condition $f'(\eta)>0$. 
Hence, the condition to be free of ghost is given by Eq.~(\ref{eq:2.A-1}). 

In case of the model in Eq.~(\ref{NLdS2}), 
for the solution in Eq.~(\ref{nlHrv9}) with $\eta_0 = \eta_1=0$ 
and in Eq.~(\ref{nlHrv10}) with $\xi_0=0$ and $\xi_1$ given 
in (\ref{nlHrv12}), 
the condition (\ref{E8}) to avoid a ghost 
is described by the following form:
\begin{equation}   
\frac{6}{\beta\left\{ 
1-\frac{3\beta}{4\left(1-6 \kappa^2 \alpha U_0^2 H_0^2 \right) -3\beta} 
+ \frac{\Lambda}{3 f_0 H_0^2 \left(1+36 \kappa^2 \alpha U_0^2 H_0^2 \right)} 
\exp \left[ \frac{4 H_0 
\left(1-6 \kappa^2 \alpha U_0^2 H_0^2 \right)}{\beta} t \right] 
\right\}} > 1\,. 
\label{eq:3.B-4} 
\end{equation}
In addition, from a necessary condition $f'(\eta)>0$ we find 
$f_0 / \beta >0$, which implies that the sign of $f_0$ is the same as 
that of $\beta$. 
In the same way, a de Sitter solution of arbitrary non-local gravity with 
Lagrange constraint multiplier may be found. 
Preliminary consideration shows that adding an $R^2$ term may also make a 
screening scenario to be realistic enough. 

\section{Finite-time future singularities in non-local gravity} 

In this section, 
we examine whether there exists the finite-time future singularities 
in non-local gravity. 
In the flat FLRW space-time, 
we analyze an asymptotic solution of 
the gravitational field equations (\ref{nl7a}) and (\ref{nl7b}) 
in the limit of the time when the finite-time future singularities 
appear.

\subsection{Finite-time future singularities}

We consider the case in which the Hubble parameter 
$H$ is expressed as 
\begin{equation} 
H \sim \frac{h_{\mathrm{s}}}{ \left( t_{\mathrm{s}} - t 
\right)^{q}}\,,
\label{eq:2.13}
\end{equation}
where 
$h_{\mathrm{s}}$ is a positive constant, 
$q$ is a non-zero constant larger than $-1$ 
$(q > -1, q \neq 0)$, and 
$t_{\mathrm{s}}$ is the time when the finite-time future singularity 
appears. 
We only consider the period $0< t < t_{\mathrm{s}}$ 
because $H$ should be real number.
When $t\to t_{\mathrm{s}}$, 
for $q>1$ and $0 < q < 1$, 
$H \sim h_{\mathrm{s}} \left( t_{\mathrm{s}} - t 
\right)^{-q}$ 
as well as 
$\dot{H} \sim q h_{\mathrm{s}} \left( t_{\mathrm{s}} - t 
\right)^{-\left(q+1\right)}$ become infinity 
and hence the scalar curvature $R$ diverges. 
For $-1 < q < 0$, $H$ is finite, 
but $\dot{H}$ becomes infinity and therefore $R$ also diverges. 

{}From Eq.~(\ref{eq:2.13}), we obtain
\begin{equation} 
a \sim a_{\mathrm{s}} \exp \left[ \frac{h_{\mathrm{s}}}{q-1} 
\left( t_{\mathrm{s}} - t 
\right)^{-\left(q-1\right)}
\right]\,, 
\label{eq:2.14}
\end{equation}
where $a_{\mathrm{s}}$ is a constant. 
By using 
$\ddot{\eta} + 3H\dot{\eta} =a^{-3} d \left( a^3 \dot{\eta} \right)/dt$ 
and Eq.~(\ref{nl8a}), 
$\eta$ is described as 
\begin{equation} 
\eta = -\int^{t} \frac{1}{a^3} 
\left( \int^{\bar{t}} 
Ra^3 d\bar{t} \right)dt\,.
\label{eq:2.15}
\end{equation}

In the limit $t\to t_{\mathrm{s}}$, 
for $q>1$, $\dot{H} \ll H^2$ and hence $R \sim 12 H^2$, 
whereas for $-1 < q < 0$ and $0 < q < 1$, 
$\dot{H} \gg H^2$ and hence $R \sim 6\dot{H}$. 
By applying these relations to Eq.~(\ref{eq:2.15}) and 
taking the leading term in terms of $\left( t_{\mathrm{s}} - t 
\right)$, we obtain 
\begin{eqnarray}  
\eta \Eqn{\sim} -\frac{4h_{\mathrm{s}}}{q-1} 
\left( t_{\mathrm{s}} - t \right)^{-\left(q-1\right)} 
+ {\eta}_{\mathrm{c}}
\quad 
(q > 1)\,,
\label{eq:2.16} \\  
\eta \Eqn{\sim} -\frac{6h_{\mathrm{s}}}{q-1} 
\left( t_{\mathrm{s}} - t \right)^{-\left(q-1\right)} 
+ {\eta}_{\mathrm{c}}
\quad 
(-1 < q < 0\,, \, 0 < q < 1)\,,
\label{eq:2.17}
\end{eqnarray}
where ${\eta}_{\mathrm{c}}$ is an integration constant. 
We describe the derivation of Eqs.~(\ref{eq:2.16}) 
and (\ref{eq:2.17}) from Eq.~(\ref{eq:2.15}) in Appendix A.

\subsection{Analysis for ${\eta}_{\mathrm{c}} \neq 0$} 

We study the case of ${\eta}_{\mathrm{c}} \neq 0$. 
(In the next subsection, we explore the case of ${\eta}_{\mathrm{c}} = 0$.) 
If the power in terms of $\left( t_{\mathrm{s}} - t \right)$ is 
negative (positive), $-\left(q-1\right) < 0$ ($>0$), 
the first term proportional to 
$\left( t_{\mathrm{s}} - t \right)^{-\left(q-1\right)}$
(the second constant term) is the leading one. 
Hence, for $q > 1$, the first term 
is the leading one, i.e., 
$\eta \propto \left( t_{\mathrm{s}} - t \right)^{-\left(q-1\right)}$, 
while for 
$-1 < q < 0$ and $0 < q < 1$, the second term 
is the leading one, i.e., $\eta \sim {\eta}_{\mathrm{c}}$. 
We note that if $q=1$ in Eq.~(\ref{eq:2.13}), it follows from 
Eq.~(\ref{nl8a}) that $\eta \sim 6h_{\mathrm{s}} \left[ 
\left( 1+2h_{\mathrm{s}}
\right)/\left( 1+3h_{\mathrm{s}} \right) \right] 
\ln \left( t_{\mathrm{s}} - t \right) + {\eta}_{\mathrm{c}}$. 

We take a form of $f(\eta)$ as 
\begin{equation} 
f(\eta) = f_{\mathrm{s}} \eta^{\sigma}\,, 
\label{eq:2.18}
\end{equation}
where $f_{\mathrm{s}} (\neq 0)$ and $\sigma (\neq 0)$ are non-zero constants. 
By using 
$\ddot{\xi} + 3H\dot{\xi} =a^{-3} d \left( a^3 \dot{\xi} \right)/dt$ 
and Eq.~(\ref{nl8b}), 
$\xi$ is written as 
\begin{equation} 
\xi = \int^{t} \frac{1}{a^3} \left( \int^{\bar{t}} 
\frac{d f(\eta)}{d \eta} 
Ra^3 d\bar{t} \right)dt\,.
\label{eq:2.19}
\end{equation}
By applying $R \sim 12 H^2$ (for $q>1$) and 
$R \sim 6\dot{H}$ (for $q<1$) 
to Eq.~(\ref{eq:2.19}) and 
taking the leading term in terms of $\left( t_{\mathrm{s}} - t 
\right)$, we acquire 
\begin{eqnarray} 
\hspace{-5mm}
\xi \Eqn{\sim} - f_{\mathrm{s}} \left(-\frac{4h_{\mathrm{s}}}{q-1} 
\right)^{\sigma} 
\left( t_{\mathrm{s}} - t \right)^{-\left(q-1\right) \sigma} 
+ {\xi}_{\mathrm{c}} 
\quad 
(q > 1)\,,
\label{eq:2.20} \\ 
\hspace{-5mm}
\xi \Eqn{\sim} 
\frac{6 f_{\mathrm{s}} h_{\mathrm{s}} \sigma 
{\eta}_{\mathrm{c}}^{\sigma -1}}{q-1} 
\left( t_{\mathrm{s}} - t \right)^{-\left(q-1\right)} 
+ {\xi}_{\mathrm{c}} 
\quad 
(-1 < q < 0\,, \, 0 < q < 1)\,,
\label{eq:2.21}
\end{eqnarray}
where ${\xi}_{\mathrm{c}}$ is an integration constant. 

If the power in terms of $\left( t_{\mathrm{s}} - t \right)$ is 
negative (positive), $-\left(q-1\right) \sigma < 0$ ($>0$), 
the first term proportional to 
$\left( t_{\mathrm{s}} - t \right)^{-\left(q-1\right) \sigma}$ 
(the second constant term) is the leading one. 
Therefore, for [$q > 1$, $\sigma > 0$], 
$\xi \propto \left( t_{\mathrm{s}} - t \right)^{-\left(q-1\right) \sigma}$, 
whereas for [$q > 1$, $\sigma < 0$] and 
[$-1 < q < 0\,, \, 0 < q < 1$], 
$\xi \sim {\xi}_{\mathrm{c}}$. 

Thus, there are three cases. 
(i) [$q > 1$, $\sigma > 0$]:\ 
$\eta \propto \left( t_{\mathrm{s}} - t \right)^{-\left(q-1\right)}$ 
and 
$\xi \propto \left( t_{\mathrm{s}} - t \right)^{-\left(q-1\right) \sigma}$. 
(ii) [$q > 1$, $\sigma < 0$]:\ 
$\eta \propto \left( t_{\mathrm{s}} - t \right)^{-\left(q-1\right)}$ 
and 
$\xi \sim {\xi}_{\mathrm{c}}$. 
(iii) [$-1 < q < 0\,, \, 0 < q < 1$]:\ 
$\eta \sim {\eta}_{\mathrm{c}}$ 
and 
$\xi \sim {\xi}_{\mathrm{c}}$. 

We examine the behavior of each term on the r.h.s. of 
Eqs.~(\ref{nl7a}) and (\ref{nl7b}) 
in the limit $t\to t_{\mathrm{s}}$, 
in particular that of the leading terms, 
and study the condition that an asymptotic solution can be obtained. 
When $t\to t_{\mathrm{s}}$, $\Lambda$, $\rho_{\mathrm{m}}$ and 
$P_{\mathrm{m}}$ can be neglected because these values are finite. 

First, we investigate each term 
on the r.h.s. of Eq.~(\ref{nl7a}). 
For case (i) [$q > 1$, $\sigma > 0$], 
the first, second and third terms evolve as 
\begin{eqnarray} 
\hspace{-15mm}
&&
-3H^2 \left(1+f(\eta) -\xi \right) 
\sim 
-3H^2 \left(f(\eta) -\xi \right) 
\sim 
-6 h_{\mathrm{s}}^2 f_{\mathrm{s}} \left(-\frac{4h_{\mathrm{s}}}{q-1} 
\right)^{\sigma} 
\left( t_{\mathrm{s}} - t \right)^{-\left[ \left(q-1\right) \sigma + 2q 
\right]}\,, 
\label{eq:2.22} \\ 
\hspace{-15mm}
&& 
\frac{1}{2} \dot{\xi} \dot{\eta} 
\sim 
-8 h_{\mathrm{s}}^2 \sigma f_{\mathrm{s}} 
\left(-\frac{4h_{\mathrm{s}}}{q-1} \right)^{\sigma-1} 
\left( t_{\mathrm{s}} - t \right)^{-\left[ \left(q-1\right) \sigma + q+1 
\right]}\,, 
\label{eq:2.23} \\ 
\hspace{-15mm}
&&
-3H \left( \frac{d f(\eta)}{d \eta} \dot{\eta} -\dot{\xi} \right) 
\sim 
24 h_{\mathrm{s}}^2 \sigma f_{\mathrm{s}} 
\left(-\frac{4h_{\mathrm{s}}}{q-1} \right)^{\sigma-1} 
\left( t_{\mathrm{s}} - t \right)^{-\left[ \left(q-1\right) \sigma + q+1 
\right]} 
\,. 
\label{eq:2.24}
\end{eqnarray} 
In the first relation in (\ref{eq:2.22}), we have used 
$1 \ll \abs{f(\eta) -\xi}$.  
In this case, 
the first term 
is proportional to 
$\left( t_{\mathrm{s}} - t \right)^{-\left[ \left(q-1\right) \sigma + 2q 
\right]}$, 
whereas 
the second and third terms are proportional to 
$\left( t_{\mathrm{s}} - t \right)^{-\left[ \left(q-1\right) \sigma + q+1 
\right]}$. 
Hence, the first term is the leading one. 
The coefficient of the first 
term cannot be zero because $h_{\mathrm{s}} \neq 0$ and 
$f_{\mathrm{s}} \neq 0$ and 
hence the leading term cannot vanish. 
This implies that in this case, 
$H \sim h_{\mathrm{s}}/\left( t_{\mathrm{s}} - t \right)^{q}$ 
in Eq.~(\ref{eq:2.13}) cannot be a solution describing 
the finite-time future singularities. 

For case (ii) [$q > 1$, $\sigma < 0$], 
the first term evolves as 
\begin{equation}  
-3H^2 \left(1+f(\eta) -\xi \right) 
\sim 
-3H^2 \left(1 -{\xi}_{\mathrm{c}} \right) 
\sim 
-3 h_{\mathrm{s}}^2 \left(1 -{\xi}_{\mathrm{c}} \right) 
\left( t_{\mathrm{s}} - t \right)^{-2q}\,, 
\label{eq:2.25} 
\end{equation}
where 
in the first relation, we have used $f(\eta) \sim 0$. 
The second and third terms are given by Eqs.~(\ref{eq:2.23}) and 
(\ref{eq:2.24}), respectively. 
In this case, 
the first term is proportional to 
$\left( t_{\mathrm{s}} - t \right)^{-2q}$, 
whereas 
the second and third terms are proportional to 
$\left( t_{\mathrm{s}} - t \right)^{-\left[ \left(q-1\right) \sigma + q+1 
\right]}$. 
Hence, the first term is the leading one. 
{}From Eq.~(\ref{eq:2.25}), we find 
the condition that the first term vanishes is given by 
\begin{equation} 
{\xi}_{\mathrm{c}} = 1\,.
\label{eq:2.26}
\end{equation}

For case (iii) [$-1 <q < 0\,, \, 0 < q < 1$], 
the first, second and third terms evolve as 
\begin{eqnarray}
&&
-3H^2 \left(1+f(\eta) -\xi \right) 
\sim 
-3 h_{\mathrm{s}}^2 \left(1 +f_{\mathrm{s}} {\eta}_{\mathrm{c}}^{\sigma} 
-{\xi}_{\mathrm{c}} \right) 
\left( t_{\mathrm{s}} - t \right)^{-2q}\,, 
\label{eq:2.27} \\ 
&& 
\frac{1}{2} \dot{\xi} \dot{\eta} 
\sim 
-18 h_{\mathrm{s}}^2 \sigma f_{\mathrm{s}} {\eta}_{\mathrm{c}}^{\sigma-1} 
\left( t_{\mathrm{s}} - t \right)^{-2q}\,, 
\label{eq:2.28} \\ 
&& 
-3H \left( \frac{d f(\eta)}{d \eta} \dot{\eta} -\dot{\xi} \right) 
\sim 
36 h_{\mathrm{s}}^2 \sigma f_{\mathrm{s}} {\eta}_{\mathrm{c}}^{\sigma-1} 
\left( t_{\mathrm{s}} - t \right)^{-2q}\,. 
\label{eq:2.29} 
\end{eqnarray} 
These terms are proportional to 
$\left( t_{\mathrm{s}} - t \right)^{-2q}$, 
which are the leading ones. 
Substituting Eqs.~(\ref{eq:2.27}), (\ref{eq:2.28}) and (\ref{eq:2.29}) into 
Eq.~(\ref{nl7a}), 
the condition that these leading terms vanish is given by 
\begin{equation} 
f_{\mathrm{s}} {\eta}_{\mathrm{c}}^{\sigma-1} 
\left( 6\sigma - {\eta}_{\mathrm{c}} 
\right) + {\xi}_{\mathrm{c}} -1 = 0\,.
\label{eq:2.31}
\end{equation}

Next, we investigate the r.h.s. of Eq.~(\ref{nl7b}). 
For case (i) [$q > 1$, $\sigma > 0$], 
the first, third and fourth terms evolve as 
\begin{eqnarray} 
&&
\left(2\dot{H} +3H^2  \right) \left(1+f(\eta) -\xi \right) 
\sim
3H^2 \left(1+f(\eta) -\xi \right) 
\sim
3H^2 \left( f(\eta) -\xi \right) 
\nonumber \\ 
&&
{}\sim 
6 h_{\mathrm{s}}^2 f_{\mathrm{s}} \left(-\frac{4h_{\mathrm{s}}}{q-1} 
\right)^{\sigma} 
\left( t_{\mathrm{s}} - t \right)^{-\left[ \left(q-1\right) \sigma + 2q 
\right]}\,, 
\label{eq:2.33} \\ 
&&
\frac{d^2 \left( f(\eta) -\xi \right)}{d t^2} 
\sim 
-8 h_{\mathrm{s}} f_{\mathrm{s}} \sigma 
\left(-\frac{4h_{\mathrm{s}}}{q-1} \right)^{\sigma -1} 
\left[ \left(q-1\right) \sigma + 1 \right]
\left( t_{\mathrm{s}} - t \right)^{-\left[ \left(q-1\right) \sigma + 2 
\right]}\,, 
\label{eq:2.34} \\ 
&&
2H \frac{d \left( f(\eta) -\xi \right)}{d t} 
\sim 
-16 h_{\mathrm{s}}^2 f_{\mathrm{s}} \sigma 
\left(-\frac{4h_{\mathrm{s}}}{q-1} \right)^{\sigma -1} 
\left( t_{\mathrm{s}} - t \right)^{-\left[ \left(q-1\right) \sigma + q+1 
\right]}\,.
\label{eq:2.35}
\end{eqnarray} 
In the first and second relations in (\ref{eq:2.33}), 
we have used $\dot{H} \ll H^2$ and $1 \ll \abs{f(\eta) -\xi}$, respectively. 
The second term is the same as the second one on the r.h.s. of 
Eq.~(\ref{nl7a}), which is given by Eq.~(\ref{eq:2.23}). 
In this case, 
the first term is proportional to 
$\left( t_{\mathrm{s}} - t \right)^{-\left[ \left(q-1\right) \sigma + 2q 
\right]}$, 
the second and fourth terms are proportional to 
$\left( t_{\mathrm{s}} - t \right)^{-\left[ \left(q-1\right) \sigma + q+1 
\right]}$, 
and the third term is proportional to 
$
\left( t_{\mathrm{s}} - t \right)^{-\left[ \left(q-1\right) \sigma + 2 
\right]}
$. 
Hence, the first term is the leading one. 
The coefficient of the first 
term cannot be zero because $h_{\mathrm{s}} \neq 0$ and 
$f_{\mathrm{s}} \neq 0$ and 
therefore in this case 
Eq.~(\ref{eq:2.13}) cannot be a solution 
for the finite-time future singularities. 

For case (ii) [$q > 1$, $\sigma < 0$], 
the first term evolves as 
\begin{equation}  
\left(2\dot{H} +3H^2  \right) \left(1+f(\eta) -\xi \right) 
\sim
3H^2 \left(1+f(\eta) -\xi \right) 
\sim
3H^2 \left(1 -{\xi}_{\mathrm{c}} \right) 
\sim 
3 h_{\mathrm{s}}^2 \left(1 -{\xi}_{\mathrm{c}} \right) 
\left( t_{\mathrm{s}} - t \right)^{-2q}\,, 
\label{eq:2.36} 
\end{equation}
where 
in the first and second relations, 
we have used $\dot{H} \ll H^2$ and $f(\eta) \sim 0$, respectively. 
The second, third, and fourth terms are given by Eqs.~(\ref{eq:2.23}), 
(\ref{eq:2.34}), and (\ref{eq:2.35}), respectively. 
In this case, 
the first term is proportional to 
$\left( t_{\mathrm{s}} - t \right)^{-2q}$, 
the second and fourth terms are proportional to 
$\left( t_{\mathrm{s}} - t \right)^{-\left[ \left(q-1\right) \sigma + q+1 
\right]}$, 
and 
the third term is proportional to 
$\left( t_{\mathrm{s}} - t \right)^{-\left[ \left(q-1\right) \sigma + 2 
\right]}$. 
Hence, the first term is the leading one. 
{}From Eq.~(\ref{eq:2.36}), we find 
the condition that the first term vanishes is given by 
$
{\xi}_{\mathrm{c}} = 1
$, 
which is equivalent to the condition in Eq.~(\ref{eq:2.26}).

For case (iii) [$-1 < q < 0\,, \, 0 < q < 1$], 
the first term, 
the former and latter parts of the third one, 
and the fourth one 
evolve as 
\begin{eqnarray}
\hspace{-15mm}
&& 
\left(2\dot{H} +3H^2  \right) \left(1+f(\eta) -\xi \right) 
\sim
2\dot{H} \left(1+f(\eta) -\xi \right) 
\nonumber \\ 
\hspace{-15mm}
&& 
\sim 
2q h_{\mathrm{s}}^2 \left(1 +f_{\mathrm{s}} {\eta}_{\mathrm{c}}^{\sigma} 
-{\xi}_{\mathrm{c}} \right) 
\left( t_{\mathrm{s}} - t \right)^{-\left( q+1 \right)}\,, 
\label{eq:2.38} \\ 
\hspace{-15mm}
&& 
\frac{d^2 f(\eta)}{d t^2} 
\sim 
36 h_{\mathrm{s}}^2 f_{\mathrm{s}} \sigma \left( \sigma -1 \right) 
{\eta}_{\mathrm{c}}^{\sigma-2} 
\left( t_{\mathrm{s}} - t \right)^{-2q} 
-6 h_{\mathrm{s}} q f_{\mathrm{s}} \sigma 
{\eta}_{\mathrm{c}}^{\sigma-1} 
\left( t_{\mathrm{s}} - t \right)^{-\left( q+1 \right)}\,, 
\label{eq:2.39} \\ 
\hspace{-15mm}
&& 
-\frac{d^2 \xi}{d t^2} 
\sim 
-6 h_{\mathrm{s}} q f_{\mathrm{s}} \sigma 
{\eta}_{\mathrm{c}}^{\sigma-1} 
\left( t_{\mathrm{s}} - t \right)^{-\left(q+1\right)}\,,
\label{eq:2.40} \\
\hspace{-15mm}
&&
2H \frac{d \left( f(\eta) -\xi \right)}{d t} 
\sim 
-24 h_{\mathrm{s}}^2 f_{\mathrm{s}} \sigma {\eta}_{\mathrm{c}}^{\sigma-1} 
\left( t_{\mathrm{s}} - t \right)^{-2q}\,. 
\label{eq:2.41} 
\end{eqnarray} 
Here, $d^2 f(\eta)/d t^2$ 
in Eq.~(\ref{eq:2.39}) and $-d^2 \xi/d t^2$ in Eq.~(\ref{eq:2.40}) are 
the former and latter parts of the third term on the r.h.s. of 
Eq.~(\ref{nl7b}). 
The second term is given by Eq.~(\ref{eq:2.28}). 
In the first relation in Eq.~(\ref{eq:2.38}), 
we have used $\dot{H} \gg H^2$. 
In this case, 
the first term, 
the second part of 
the former part of the third one given by Eq.~(\ref{eq:2.39}), 
and the latter part of the third one given by Eq.~(\ref{eq:2.40}) 
are proportional to 
$\left( t_{\mathrm{s}} - t \right)^{-\left( q+1 \right)}$, 
whereas 
the second term, 
the first part of 
the former part of the third one given by Eq.~(\ref{eq:2.39}), 
and the fourth one 
are proportional to 
$\left( t_{\mathrm{s}} - t \right)^{-2q}$. 
The terms proportional to 
$\left( t_{\mathrm{s}} - t \right)^{-\left( q+1 \right)}$ 
are the leading ones. 
{}From Eqs.~(\ref{eq:2.38}), (\ref{eq:2.39}) and (\ref{eq:2.40}), we find 
the condition that the leading terms vanish is given by 
$
f_{\mathrm{s}} {\eta}_{\mathrm{c}}^{\sigma-1} 
\left( 6\sigma - {\eta}_{\mathrm{c}} 
\right) + {\xi}_{\mathrm{c}} -1 = 0
$, 
which is the same as the one in Eq.~(\ref{eq:2.31}).

As a consequence, 
for case (ii) [$q > 1$, $\sigma < 0$], 
if the condition ${\xi}_{\mathrm{c}} = 1$ 
in Eq.~(\ref{eq:2.26}) is satisfied, 
the leading term vanishes in both Eqs.~(\ref{nl7a}) and (\ref{nl7b}). 
Furthermore, 
for case (iii) [$-1 < q < 0\,, \, 0 < q < 1$], 
when the condition 
$
f_{\mathrm{s}} {\eta}_{\mathrm{c}}^{\sigma-1} 
\left( 6\sigma - {\eta}_{\mathrm{c}} 
\right) + {\xi}_{\mathrm{c}} -1 = 0
$ in Eq.~(\ref{eq:2.31}) 
is met, 
the leading terms also vanish 
in both Eqs.~(\ref{nl7a}) and (\ref{nl7b}). 
Thus, 
the expression of 
the Hubble parameter $H$ in Eq.~(\ref{eq:2.13}) 
can be a leading-order solution in terms of 
$\left( t_{\mathrm{s}} - t \right)$ 
for the gravitational field equations 
in the flat FLRW space-time. 
This implies that there can exist 
the finite-time future singularities 
in non-local gravity. 

\subsection{Analysis for ${\eta}_{\mathrm{c}}= 0$} 

If we take ${\eta}_{\mathrm{c}} = 0$ in Eqs.~(\ref{eq:2.16}) and 
(\ref{eq:2.17}), by using Eqs.~(\ref{eq:2.19}) we see that 
for $q>1$, $\xi$ is given by Eq.~(\ref{eq:2.20}), 
whereas for $q<1$, $\xi$ is described as 
\begin{equation} 
\xi \sim 
- \frac{q f_{\mathrm{s}}}{\left(q-1\right) \sigma + 1} 
\left(-\frac{6h_{\mathrm{s}}}{q-1} \right)^{\sigma} 
\left( t_{\mathrm{s}} - t \right)^{-\left(q-1\right) \sigma} 
+ {\xi}_{\mathrm{c}} 
\quad 
(-1 < q < 0\,, \, 0 < q < 1)\,. 
\label{eq:2.44}
\end{equation}
If the power in terms of $\left( t_{\mathrm{s}} - t \right)$ is 
negative (positive), $-\left(q-1\right) \sigma < 0$ ($>0$), 
the first term proportional to 
$\left( t_{\mathrm{s}} - t \right)^{-\left(q-1\right) \sigma}$ 
(the second constant term) is the leading one. 
Therefore, 
for [$-1 < q < 0\,, \, 0 < q < 1$, $\sigma > 0$], 
$\xi \sim {\xi}_{\mathrm{c}}$, 
while for 
[$-1 < q < 0\,, \, 0 < q < 1$, $\sigma < 0$], 
$\xi \propto \left( t_{\mathrm{s}} - t \right)^{-\left(q-1\right) \sigma}$.  

Thus, case (iii) is divided into two cases. 
One is 
case (iii-a) [$-1 < q < 0\,, \, 0 < q < 1$, $\sigma > 0$]:\ 
$\eta \propto \left( t_{\mathrm{s}} - t \right)^{-\left(q-1\right)}$ 
and 
$\xi \sim {\xi}_{\mathrm{c}}$. 
Another is 
case (iii-b) [$-1 < q < 0\,, \, 0 < q < 1$, $\sigma < 0$]:\ 
$\eta \propto \left( t_{\mathrm{s}} - t \right)^{-\left(q-1\right)}$ 
and 
$\xi \propto \left( t_{\mathrm{s}} - t \right)^{-\left(q-1\right) \sigma}$. 

First, we investigate each term 
on the r.h.s. of Eq.~(\ref{nl7a}). 
For case (iii-a) [$-1 < q < 0\,, \, 0 < q < 1$, $\sigma > 0$], 
the first, second and third terms 
evolve as 
\begin{eqnarray}
\hspace{-15mm}
&&
-3H^2 \left(1+f(\eta) -\xi \right) 
\sim 
-3H^2 f(\eta) 
\sim 
-3 h_{\mathrm{s}}^2 f_{\mathrm{s}} 
\left(-\frac{6h_{\mathrm{s}}}{q-1} \right)^{\sigma} 
\left( t_{\mathrm{s}} - t \right)^{-\left[ \left(q-1\right) \sigma + 2q 
\right]}\,,  
\label{eq:2.45} \\ 
\hspace{-15mm}
&& 
\frac{1}{2} \dot{\xi} \dot{\eta} 
\sim 
-18 h_{\mathrm{s}}^2 \sigma f_{\mathrm{s}} 
\left(-\frac{6h_{\mathrm{s}}}{q-1} \right)^{\sigma-1} 
\frac{q}{\left( q-1 \right)\sigma +1} 
\left( t_{\mathrm{s}} - t \right)^{-\left[ \left(q-1\right) \sigma + q+1 
\right]}\,, 
\label{eq:2.46} \\ 
\hspace{-15mm}
&&
-3H \left( \frac{d f(\eta)}{d \eta} \dot{\eta} -\dot{\xi} \right) 
\sim 
18 h_{\mathrm{s}}^2 \sigma f_{\mathrm{s}} 
\left(-\frac{6h_{\mathrm{s}}}{q-1} \right)^{\sigma-1} 
\frac{\left( q-1 \right)\sigma +q+1}{\left( q-1 \right)\sigma +1} 
\left( t_{\mathrm{s}} - t \right)^{-\left[ \left(q-1\right) \sigma + q+1 
\right]}\,. 
\label{eq:2.47}
\end{eqnarray} 
In the first relation in (\ref{eq:2.45}), we have used 
$\abs{1 - {\xi}_{\mathrm{c}}} \ll \abs{f(\eta)}$. 
In this case, 
the first term 
is proportional to 
$\left( t_{\mathrm{s}} - t \right)^{-\left[ \left(q-1\right) \sigma + 2q 
\right]}$, 
whereas 
the second and third terms are proportional to 
$\left( t_{\mathrm{s}} - t \right)^{-\left[ \left(q-1\right) \sigma + q+1 
\right]}$. 
Hence, the second and third terms are the leading ones 
and 
the coefficient of these leading terms is given by 
$
18 h_{\mathrm{s}}^2 \sigma f_{\mathrm{s}} 
\left[ -6h_{\mathrm{s}}/\left(q-1\right) \right]^{\sigma-1} 
$. 
This cannot be zero because $h_{\mathrm{s}} \neq 0$, 
$f_{\mathrm{s}} \neq 0$ and $\sigma \neq 0$ 
and hence the leading terms cannot vanish. 
This implies that in this case, 
$H \sim h_{\mathrm{s}}/\left( t_{\mathrm{s}} - t \right)^{q}$ 
in Eq.~(\ref{eq:2.13}) cannot be a solution describing 
the finite-time future singularities. 

For case (iii-b) [$-1 < q < 0\,, \, 0 < q < 1$, $\sigma < 0$], 
the first term evolves as 
\begin{eqnarray}
&&
-3H^2 \left(1+f(\eta) -\xi \right) 
\sim 
-3H^2 \left(f(\eta) -\xi \right) 
\nonumber \\
&&
\sim 
-3 h_{\mathrm{s}}^2 f_{\mathrm{s}} 
\left(-\frac{6h_{\mathrm{s}}}{q-1} \right)^{\sigma} 
\frac{\left( q-1 \right)\sigma +q+1}{\left( q-1 \right)\sigma +1} 
\left( t_{\mathrm{s}} - t \right)^{-\left[ \left(q-1\right) \sigma + 2q 
\right]}\,, 
\label{eq:2.48} 
\end{eqnarray} 
where in the first relation, we have used $1 \ll \abs{f(\eta) -\xi}$.  
The second and third terms are given by Eqs.~(\ref{eq:2.46}) and 
(\ref{eq:2.47}), respectively. 
Since the first term 
is proportional to 
$\left( t_{\mathrm{s}} - t \right)^{-\left[ \left(q-1\right) \sigma + 2q 
\right]}$, 
which is the same as in Eq.~(\ref{eq:2.45}), 
the second and third terms are the leading ones. 
Thus, the consequence is equivalent to that in case (iii-a). 

Next, we investigate the r.h.s. of Eq.~(\ref{nl7b}). 
For case (iii-a) [$-1 < q < 0\,, \, 0 < q < 1$, $\sigma > 0$], 
the first, third and fourth terms evolve as 
\begin{eqnarray}
\hspace{-15mm}
&&
\left(2\dot{H} +3H^2  \right) \left(1+f(\eta) -\xi \right) 
\sim
2\dot{H} \left(1+f(\eta) -\xi \right) 
\sim 
-2\dot{H} f(\eta) 
\nonumber \\ 
\hspace{-15mm}
&&
\sim 
2 h_{\mathrm{s}} f_{\mathrm{s}} 
\left(-\frac{6h_{\mathrm{s}}}{q-1} \right)^{\sigma} 
\left( t_{\mathrm{s}} - t \right)^{-\left[ \left(q-1\right) \sigma +q+1 
\right]}\,, 
\label{eq:2.49} \\
\hspace{-15mm}
&&
\frac{d^2 \left( f(\eta) -\xi \right)}{d t^2} 
\sim 
36 h_{\mathrm{s}}^2 f_{\mathrm{s}} \sigma 
\left(-\frac{6h_{\mathrm{s}}}{q-1} \right)^{\sigma -2} 
\frac{\left( q-1 \right)\sigma +q+1}{q-1} 
\left( t_{\mathrm{s}} - t \right)^{-\left[ \left(q-1\right) \sigma + 2 
\right]}\,, 
\label{eq:2.50} \\ 
\hspace{-15mm}
&&
2H \frac{d \left( f(\eta) -\xi \right)}{d t} 
\sim 
-12 h_{\mathrm{s}}^2 f_{\mathrm{s}} \sigma 
\left(-\frac{6h_{\mathrm{s}}}{q-1} \right)^{\sigma -1} 
\frac{\left( q-1 \right)\sigma +q+1}{\left( q-1 \right)\sigma +1} 
\left( t_{\mathrm{s}} - t \right)^{-\left[ \left(q-1\right) \sigma + q+1 
\right]}\,.
\label{eq:2.51}
\end{eqnarray} 
In the first and second relations in (\ref{eq:2.49}), 
we have used $\dot{H} \gg H^2$ and 
$\abs{1 - {\xi}_{\mathrm{c}}} \ll \abs{f(\eta)}$, respectively. 
The second term is the same as the second one on the r.h.s. of 
Eq.~(\ref{nl7a}), which is given by Eq.~(\ref{eq:2.46}). 
In this case, 
the first, second and fourth terms 
are proportional to 
$\left( t_{\mathrm{s}} - t \right)^{-\left[ \left(q-1\right) \sigma + q+1 
\right]}$, 
whereas 
the third term 
$\left( t_{\mathrm{s}} - t \right)^{-\left[ \left(q-1\right) \sigma + 2 
\right]}$. 
Hence, the third term is the leading one. 
{}From Eq.~(\ref{eq:2.50}), we find 
the condition that the third term vanishes is given by 
\begin{equation} 
\sigma = -\frac{q+1}{q-1}\,. 
\label{eq:2.52}
\end{equation}

For case (iii-b) [$-1 < q < 0\,, \, 0 < q < 1$, $\sigma < 0$], 
the first term evolves as 
\begin{eqnarray} 
&&
\left(2\dot{H} +3H^2  \right) \left(1+f(\eta) -\xi \right) 
\sim
2\dot{H} \left(1+f(\eta) -\xi \right) 
\sim
2\dot{H} \left( f(\eta) -\xi \right) 
\nonumber \\ 
&&
{}\sim 
2q h_{\mathrm{s}} f_{\mathrm{s}} 
\left(-\frac{6h_{\mathrm{s}}}{q-1} \right)^{\sigma} 
\frac{\left( q-1 \right)\sigma +q+1}{\left( q-1 \right)\sigma +1} 
\left( t_{\mathrm{s}} - t \right)^{-\left[ \left(q-1\right) \sigma + q+1 
\right]}\,, 
\label{eq:2.53}  
\end{eqnarray} 
where 
in the first and second relation, we have used 
$\dot{H} \gg H^2$ and 
$1 \ll \abs{f(\eta) -\xi}$, respectively. 
The second, third, and fourth terms are given by Eqs.~(\ref{eq:2.46}), 
(\ref{eq:2.50}) and (\ref{eq:2.51}), respectively. 
Since the first term 
is proportional to 
$\left( t_{\mathrm{s}} - t \right)^{-\left[ \left(q-1\right) \sigma + q+1 
\right]}$, 
which is the same as in Eq.~(\ref{eq:2.49}), 
the third term is the leading one. 
Thus, the consequence is equivalent to that in case (iii-a). 

As a result, 
if we take ${\eta}_{\mathrm{c}} = 0$, 
we cannot have the case in which 
the leading term vanishes 
in both Eqs.~(\ref{nl7a}) and (\ref{nl7b}).

\subsection{Relations between the model parameters and the property of 
the finite-time future singularities} 

The parameters $f_{\mathrm{s}}$ and $\sigma$ in the form of $f(\eta)$ 
in Eq.~(\ref{eq:2.18}) characterize the theory of non-local gravity 
described by the action in Eq.~(\ref{nl1}). 
On the other hand, 
the parameters 
$h_{\mathrm{s}}$, $t_{\mathrm{s}}$ and $q$ 
in the expression of the Hubble parameter $H$ in Eq.~(\ref{eq:2.13}) 
specify the property of the finite-time future singularity. 
Moreover, the integration constants 
${\eta}_{\mathrm{c}}$ in Eqs.~(\ref{eq:2.16}) and (\ref{eq:2.17}) 
and ${\xi}_{\mathrm{c}}$ in Eqs.~(\ref{eq:2.20}) and (\ref{eq:2.21}) 
determine a leading-order solution in terms of 
$\left( t_{\mathrm{s}} - t \right)$ 
for the gravitational field equations 
in the flat FLRW space-time. 

{}From Eq.~(\ref{eq:2.14}), we see that 
when $t\to t_{\mathrm{s}}$, 
for $q>1$, $a \to \infty$, 
whereas 
for $-1 < q < 0$ and $0 < q < 1$, 
$a \to a_{\mathrm{s}}$. 
Moreover, 
it follows from Eqs.~(\ref{eq:2.13}) and (\ref{eq:2.C-1}) that 
for $q>0$, $H \to \infty$ and therefore 
$\rho_{\mathrm{eff}} = 3 H^2/\kappa^2 \to \infty$, 
whereas 
for $-1 < q < 0$, $H$ asymptotically becomes finite 
and also $\rho_{\mathrm{eff}}$ asymptotically approaches a finite 
constant value $\rho_{\mathrm{s}}$. 
On the other hand, from 
$\dot{H} \sim q h_{\mathrm{s}} \left( t_{\mathrm{s}} - t 
\right)^{-\left(q+1\right)}$ and Eq.~(\ref{eq:2.C-2}) 
we find that 
for $q>-1$, $\dot{H} \to \infty$ and hence 
$P_{\mathrm{eff}} = -\left(2\dot H + 3H^2\right)/\kappa^2 \to \infty$. 

It is known that 
the finite-time future singularities\footnote{
For the first observation of the finite-time future singularities 
in $F(R)$ gravity, see~\cite{FS-F(R)-gravity}. 
In Ref.~\cite{Future-singularity-MG}, 
the finite-time future singularities in various modified gravity 
theories have been examined.} can be classified 
in the following manner~\cite{Nojiri:2005sx}: 
\begin{itemize}
\item Type I (``Big Rip''~\cite{Big-Rip}):\ 
In the limit $t\to t_{\mathrm{s}}$, 
$a \to \infty$,
$\rho_{\mathrm{eff}} \to \infty$ and
$\abs{ P_{\mathrm{eff}} } \to \infty$. 
The case in which 
$\rho_\mathrm{{eff}}$ and $P_{\mathrm{eff}}$ becomes finite values 
at $t = t_{\mathrm{s}}$~\cite{Shtanov:2002ek} 
is also included. 
\item Type II (``sudden''~\cite{sudden, Koivisto:2008xfa}):\ 
In the limit $t\to t_{\mathrm{s}}$, 
$a \to a_{\mathrm{s}}$, 
$\rho_{\mathrm{eff}} \to \rho_{\mathrm{s}}$ and 
$\abs{ P_{\mathrm{eff}} } \to \infty$. 
\item Type III:\ 
In the limit $t\to t_{\mathrm{s}}$, 
$a \to a_{\mathrm{s}}$, 
$\rho_{\mathrm{eff}} \to \infty$ and
$\abs{ P_{\mathrm{eff}} } \to \infty$. 
\item Type IV:\ 
In the limit $t\to t_{\mathrm{s}}$, 
$a \to a_{\mathrm{s}}$, 
$\rho_{\mathrm{eff}} \to 0$, 
$\abs{ P_{\mathrm{eff}} } \to 0$, 
and higher derivatives of $H$ diverge. 
The case in which $\rho_{\mathrm{eff}}$ and/or $\abs{ P_{\mathrm{eff}} }$ 
asymptotically approach finite values is also included. 
\end{itemize}
%

As a consequence, the finite-time future singularities 
described by the expression of $H$ 
in Eq.~(\ref{eq:2.13}) in non-local gravity 
have the following properties: 
For $q>1$, the Type I (``Big Rip'') singularity, 
for $0<q<1$, the Type III singularity, 
and for $-1<q<0$, the Type II (``sudden'') singularity. 

The results obtained in Secs. IV A and IV B 
are summarized in Table \ref{tb:table1}. 
In Table \ref{tb:table1}, 
we show the range and conditions for the value of parameters 
of $f(\eta)$, $H$, and ${\eta}_{\mathrm{c}}$ and ${\xi}_{\mathrm{c}}$ 
in order that the finite-time future singularities 
can exist. 
If ${\eta}_{\mathrm{c}} \neq 0$ and ${\xi}_{\mathrm{c}} = 1$, 
in a model with $\sigma < 0$, 
there can exist the finite-time future singularities 
with the property of 
the Type I (``Big Rip'') singularity for $q > 1$. 
If ${\eta}_{\mathrm{c}} \neq 0$, 
in a model with satisfying the condition 
$f_{\mathrm{s}} {\eta}_{\mathrm{c}}^{\sigma-1} 
\left( 6\sigma - {\eta}_{\mathrm{c}} 
\right) + {\xi}_{\mathrm{c}} -1 = 0$, 
there can exist the finite-time future singularities 
with the property of 
the Type III singularity for $0 < q < 1$ 
and that of the Type II (``sudden'') singularity for $-1 < q < 0$. 
We remark that for the special case of ${\eta}_{\mathrm{c}}= 0$, 
the finite-time future singularities 
described by $H$ in Eq.~(\ref{eq:2.13}) cannot occur. 

\begin{table*}[tbp]
\caption{Range and conditions for the value of parameters 
of $f(\eta)$, $H$, and ${\eta}_{\mathrm{c}}$ and ${\xi}_{\mathrm{c}}$ 
in order that the finite-time future singularities 
can exist. 
}
\begin{center}
\begin{tabular}
{llll}
\hline
\hline
Case
& $f(\eta) = f_{\mathrm{s}} \eta^{\sigma}$
& $H \sim \frac{h_{\mathrm{s}}}{ \left( t_{\mathrm{s}} - t 
\right)^{q}}$
& ${\eta}_{\mathrm{c}}\,, \,\, {\xi}_{\mathrm{c}}$
\\[0mm]
\hline
&
$f_{\mathrm{s}} \neq 0$
& 
$h_{\mathrm{s}} > 0$
& 
${\eta}_{\mathrm{c}} \neq 0$
\\[0mm]
&
$\sigma \neq 0$
&
$q > -1\,, \,\, q \neq 0$
&
\\[0mm]
(ii)
&$\sigma < 0$
&$q > 1$ [Type I (``Big Rip'') singularity] 
&${\xi}_{\mathrm{c}} = 1$
\\[0mm]
(iii) \quad {}
&$f_{\mathrm{s}} {\eta}_{\mathrm{c}}^{\sigma-1} 
\left( 6\sigma - {\eta}_{\mathrm{c}} 
\right) + {\xi}_{\mathrm{c}} -1 = 0$
\quad {}
&$0 < q < 1$ [Type III singularity]
\quad {}
&
\\[0mm]
& 
&$-1 < q < 0$ [Type II (``sudden'') singularity]
& 
\\[1mm]
\hline
\hline
\end{tabular}
\end{center}
\label{tb:table1}
\end{table*}

\section{Effective equation of state for the universe and 
phantom-divide crossing}

\subsection{Cosmological evolution of the effective equation of state for 
the universe}

We study the effective equation of state for the universe 
when the finite-time future singularities 
occur in non-local gravity. 
We examine the asymptotic behavior of $w_{\mathrm{eff}}$ 
given by Eq.~(\ref{eq:5.1}) 
in the limit $t\to t_{\mathrm{s}}$ by using the 
results obtained in Sec.~IV B. 
First, we study 
case (ii) [$q > 1$, $\sigma < 0$] [Type I (``Big Rip'') singularity]. 
The first term in the denominator of Eq.~(\ref{eq:5.1}) is given by 
$-3H^2 \left( f(\eta) -\xi \right) 
\sim 3 h_{\mathrm{s}}^2 \left( t_{\mathrm{s}} - t \right)^{-2q}$, 
where we have used $f(\eta) \sim 0$ and Eq.~(\ref{eq:2.26}). 
Moreover, the first term in the numerator of Eq.~(\ref{eq:5.1}) is given by 
$\left(2\dot{H} +3H^2  \right) \left( f(\eta) -\xi \right) 
\sim - 3 h_{\mathrm{s}}^2 \left( t_{\mathrm{s}} - t \right)^{-2q}$, 
where we have used 
$\dot{H} \ll H^2$, $f(\eta) \sim 0$ and ${\xi}_{\mathrm{c}} = 1$. 
We note that these first terms proportional to 
$\left( t_{\mathrm{s}} - t \right)^{-2q}$ are the leading ones 
in the denominator and numerator of Eq.~(\ref{eq:5.1}) and 
therefore $\Lambda$, $\kappa^2 \rho_{\mathrm{m}}$ and 
$\kappa^2 P_{\mathrm{m}}$ can be neglected. 
By substituting these relations, 
Eqs.~(\ref{eq:2.23}), (\ref{eq:2.24}), (\ref{eq:2.34}) and (\ref{eq:2.35}) 
into Eq.~(\ref{eq:5.1}) 
and taking the leading term in terms of $\left( t_{\mathrm{s}} - t \right)$, 
we find 
\begin{equation} 
w_{\mathrm{eff}} \sim 
-1 + I(t) 
\sim -1\,, 
\label{eq:5.4}
\end{equation}
where 
\begin{equation}   
I(t) = 
-8 \sigma f_{\mathrm{s}} 
\left(-\frac{4h_{\mathrm{s}}}{q-1} \right)^{\sigma-1} 
\left( t_{\mathrm{s}} - t \right)^{\left(q-1\right)\left(1-\sigma\right)}\,. 
\label{eq:5.5} 
\end{equation}
Here, $I(t)$ corresponds to the deviation of $w_{\mathrm{eff}}$ from $-1$. 
In deriving the second relation in (\ref{eq:5.4}), we have used the fact that 
$\left( t_{\mathrm{s}} - t \right)^{\left(q-1\right)\left(1-\sigma\right)} 
\sim 0$ 
because the power in terms of $\left( t_{\mathrm{s}} - t \right)$ is 
positive ($\left(q-1\right)\left(1-\sigma\right) > 0$). 
Since 
$h_{\mathrm{s}} >0$, 
$q > 1$ and $\sigma < 0$, from (\ref{eq:5.4}) and Eq.~(\ref{eq:5.5}) 
we see that 
if $(-)^{\sigma-1} f_{\mathrm{s}} >0$, 
$I(t)$ evolves from $I(t) > 0$ to $I(t) = 0$, 
and hence 
$w_\mathrm{eff}$ evolves from 
$w_\mathrm{eff} > -1$: the non-phantom (quintessence) phase 
and asymptotically approaches $w_\mathrm{eff} = -1$ 
corresponding to the phase of the cosmological constant, 
whereas 
if $(-)^{\sigma-1} f_{\mathrm{s}} <0$, 
$I(t)$ evolves from $I(t) < 0$ to $I(t) = 0$, 
and therefore 
$w_\mathrm{eff}$ evolves from 
$w_\mathrm{eff} < -1$: the phantom phase 
and asymptotically approaches 
$w_\mathrm{eff} = -1$.

Next, we explore case (iii) [$-1 < q < 0\,, \, 0 < q < 1$]. 
The first term in the denominator of Eq.~(\ref{eq:5.1}) is given by 
$-3H^2 \left( f(\eta) -\xi \right) 
\sim -3 h_{\mathrm{s}}^2 \left( f_{\mathrm{s}} {\eta}_{\mathrm{c}}^{\sigma} 
-{\xi}_{\mathrm{c}} \right) \left( t_{\mathrm{s}} - t \right)^{-2q}$. 
Moreover, the first term in the numerator of Eq.~(\ref{eq:5.1}) is given by 
$\left(2\dot{H} +3H^2  \right) \left( f(\eta) -\xi \right) 
\sim 2q h_{\mathrm{s}}^2 \left(1 +f_{\mathrm{s}} {\eta}_{\mathrm{c}}^{\sigma} 
-{\xi}_{\mathrm{c}} \right) 
\left( t_{\mathrm{s}} - t \right)^{-\left( q+1 \right)}$, 
where we have used $\dot{H} \gg H^2$. 
By substituting these relations, 
Eqs.~(\ref{eq:2.28}), (\ref{eq:2.29}), (\ref{eq:2.39}), 
(\ref{eq:2.40}) and (\ref{eq:2.41}) into Eq.~(\ref{eq:5.1}) 
and using the condition in Eq.~(\ref{eq:2.31}), 
we obtain the expression of Eq.~(\ref{eq:5.1}) and 
find that in the denominator of Eq.~(\ref{eq:5.1}) 
there is a term proportional to 
$\left( t_{\mathrm{s}} - t \right)^{-2q}$ and 
in the numerator 
there are a term proportional to 
$\left( t_{\mathrm{s}} - t \right)^{-2q}$ 
and that proportional to 
$\left( t_{\mathrm{s}} - t \right)^{-\left( q+1 \right)}$. 
If the power of the terms proportional to 
$\left( t_{\mathrm{s}} - t \right)^{-2q}$
is negative, i.e., for $0 < q < 1$ [Type III singularity], 
the power of the term proportional to 
$\left( t_{\mathrm{s}} - t \right)^{-\left( q+1 \right)}$ 
is also negative, 
and therefore 
in the limit $t\to t_{\mathrm{s}}$ 
the terms proportional to 
$\left( t_{\mathrm{s}} - t \right)^{-2q}$ 
and 
that proportional to 
$\left( t_{\mathrm{s}} - t \right)^{-\left( q+1 \right)}$ 
become dominant over 
($\Lambda 
+ \kappa^2 \rho_{\mathrm{m}}$) and ($- \Lambda + \kappa^2 P_{\mathrm{m}}$). 
Hence, $\Lambda$, $\kappa^2 \rho_{\mathrm{m}}$ and $\kappa^2 P_{\mathrm{m}}$ 
can be neglected. 
As a consequence, Eq.~(\ref{eq:5.1}) is described as 
\begin{equation}   
w_{\mathrm{eff}} \sim 
-1 + I(t) 
\sim 
-\frac{2q}{3h_{\mathrm{s}}} 
\left( t_{\mathrm{s}} - t \right)^{q-1}\,,  
\label{eq:5.6} 
\end{equation}
where
\begin{eqnarray} 
I(t) \Eqn{=} I_0 
-\frac{2q}{3h_{\mathrm{s}}} 
\left( t_{\mathrm{s}} - t \right)^{q-1}\,, 
\label{eq:5.7} \\
I_0 \Eqn{=} 
1+ 2 f_{\mathrm{s}} \sigma {\eta}_{\mathrm{c}}^{\sigma-2} 
\left[ 6\left( \sigma -1 \right) -7{\eta}_{\mathrm{c}} \right]\,. 
\label{eq:5.8} 
\end{eqnarray} 
Here, $I_0$ is a constant part of $I(t)$. 
In deriving the second relation in (\ref{eq:5.6}), 
we have used the fact that 
in the limit $t\to t_{\mathrm{s}}$, the second term 
in Eq.~(\ref{eq:5.7}) 
becomes dominant over the first one 
because the power in terms of $\left( t_{\mathrm{s}} - t \right)$ is 
negative ($q-1 < 0$). 
If $I_0 > 0$, 
a crossing of the phantom 
divide from the non-phantom 
phase to the phantom one can occur 
because the sign of the 
second term in Eq.~(\ref{eq:5.7}) is negative and 
the absolute value of the amplitude becomes very large, 
while if $I_0 < 0$, 
$I(t)$ always evolves in the phantom phase 
($w_\mathrm{eff} < -1$). 
Thus, for $0 < q < 1$ 
the final stage is the phantom phase 
and it is eternal. 
It seems that the recent cosmological observational data imply 
the crossing of the phantom divide 
from the non-phantom phase to the phantom one 
in the near past~\cite{observational status}. 
The (past or) future crossing of the phantom divide in $F(R)$ gravity have 
also been investigated in Ref.~\cite{Future-Crossing-in-F(R)}. 

On the other hand,
if $-1 < q < 0$ [Type II (``sudden'') singularity], 
in the denominator of Eq.~(\ref{eq:5.1}) 
the term proportional to 
$\left( t_{\mathrm{s}} - t \right)^{-2q}$ 
asymptotically vanishes and ($\Lambda 
+ \kappa^2 \rho_{\mathrm{m}}$) become dominant, 
whereas in the numerator 
the term proportional to 
$\left( t_{\mathrm{s}} - t \right)^{-\left( q+1 \right)}$ 
again becomes dominant over ($- \Lambda + \kappa^2 P_{\mathrm{m}}$), 
and thus $\Lambda$ and $\kappa^2 P_{\mathrm{m}}$ can be neglected. 
As a result, 
we acquire 
\begin{equation}  
w_{\mathrm{eff}} \sim 
-2 h_{\mathrm{s}} q 
\left( t_{\mathrm{s}} - t \right)^{-\left( q+1 \right)}/\left( 
\Lambda + \kappa^2 \rho_{\mathrm{m}} \right)\,. 
\label{eq:5.9} 
\end{equation}
The absolute value of the amplitude of the term 
$-2 h_{\mathrm{s}} q 
\left( t_{\mathrm{s}} - t \right)^{-\left( q+1 \right)}$ 
becomes very large. 
If we consider $\Lambda > 0$, 
since $h_{\mathrm{s}} > 0$ and $-1 < q < 0$, 
we have $w_{\mathrm{eff}} > 0$. 

In summary, 
for $q > 1$ [Type I (``Big Rip'') singularity], 
$w_{\mathrm{eff}}$ evolves from 
the non-phantom phase or the phantom one 
and asymptotically approaches $w_\mathrm{eff} = -1$.  
For $0 < q < 1$ [Type III singularity], 
$w_{\mathrm{eff}}$ evolves from the non-phantom 
to the phantom one with realizing 
a crossing of the phantom divide 
or evolves in the phantom phase. 
The final stage is the eternal phantom phase. 
For $-1 < q < 0$ [Type II (``sudden'') singularity], 
$w_{\mathrm{eff}} > 0$ at the final stage.

\subsection{Estimation of the current value of the effective equation of state 
parameter for non-local gravity}

In Ref.~\cite{Komatsu:2010fb}, 
the limit on a constant equation of state for dark energy 
in a flat universe has been estimated as 
$w_{\mathrm{DE}} = -1.10 \pm 0.14 \, 
(68 \% \, \mathrm{CL})$ 
by combining the data of 
Seven-Year Wilkinson Microwave Anisotropy Probe (WMAP) 
Observations~\cite{Komatsu:2010fb} 
with the latest distance measurements from the BAO in the 
distribution of galaxies and the Hubble constant measurement. 
Moreover, for a time-dependent equation of state for dark energy, 
by using a linear form $w_{\mathrm{DE}}(a) = w_{\mathrm{DE}\,0} + 
w_{\mathrm{DE}\,a} \left( 1-a \right)$~\cite{C-P-L} 
with $w_{\mathrm{DE}\,0}$ and $w_{\mathrm{DE}\,a}$ 
being the current value of $w_{\mathrm{DE}}$ and 
its derivative, 
constraints on $w_{\mathrm{DE}\,0}$ and $w_{\mathrm{DE}\,a}$ 
have been found as 
$w_{\mathrm{DE}\,0} = -0.93 \pm 0.13$ and 
$w_{\mathrm{DE}\,a} = -0.41^{+0.72}_{-0.71} \, (68 \% \, \mathrm{CL})$, 
from the combination of the WMAP data with the BAO data, 
the Hubble constant measurement and the high-redshift SNe Ia data. 

We estimate the present value of $w_{\mathrm{eff}}$. 
Here, 
we regard $w_{\mathrm{eff}}$ as being approximately equal to $w_{\mathrm{DE}}$ 
at the present time ($w_{\mathrm{eff}} \approx w_{\mathrm{DE}}$) 
because the energy density of dark energy is dominant over that of 
non-relativistic matter at the present time. 
For case (ii) [$q > 1$, $\sigma < 0$], 
we take 
$\sigma = -1$, $q = 2$, $h_{\mathrm{s}} = 1 \, [\mathrm{GeV}]^{-1}$ 
and $t_{\mathrm{s}} = 2 t_{\mathrm{p}}$, 
where $t_{\mathrm{p}}$ is the present time. 
The current value of the Hubble parameter is given by  
$H_{\mathrm{p}} = 2.1 h \times 10^{-42} \mathrm{GeV}$~\cite{Kolb and Turner}
with $h = 0.7$~\cite{Komatsu:2010fb, Freedman:2000cf}. 
We note that in Eq.~(\ref{eq:2.13}), $h_{\mathrm{s}}$ has the dimension of 
$[\mathrm{Mass}]^{q-1}$ and $t_{\mathrm{s}}$ and $t$ have the dimension of 
$[\mathrm{Mass}]^{-1}$. 
In this case, by using the first approximate relation in (\ref{eq:5.4}) and 
(\ref{eq:5.5}), we find that 
if $f_{\mathrm{s}} = -3.0 \times 10^{-43}$, $w_{\mathrm{eff}} = -1.10$, and 
if $f_{\mathrm{s}} = -2.1 \times 10^{-43}$, $w_{\mathrm{eff}} = -0.93$. 
For $0 < q < 1$, 
we take 
$\sigma = 1$, $q = 1/2$, $h_{\mathrm{s}} = 1 \, [\mathrm{GeV}]^{1/2}$, 
${\eta}_{\mathrm{c}} = 1$ and $t_{\mathrm{s}} = 2 t_{\mathrm{p}}$. 
In this case, by using the first approximate relation in (\ref{eq:5.6}) and 
Eqs.~(\ref{eq:5.7}) and (\ref{eq:5.8}), we see that 
if $f_{\mathrm{s}} = 7.9 \times 10^{-2}$, $w_{\mathrm{eff}} = -1.10$, and 
if $f_{\mathrm{s}} = 6.6 \times 10^{-2}$, $w_{\mathrm{eff}} = -0.93$. 
For $-1 < q < 0$, from Eq.~(\ref{eq:5.9}) we have $w_{\mathrm{eff}} > 0$. 
Thus, 
in our models 
$w_{\mathrm{eff}}$ can have 
the present observed value of $w_{\mathrm{DE}}$ estimated in 
Ref.~\cite{Komatsu:2010fb}. 

We remark that in this subsection, 
as a demonstration we have examined whether our model can 
have the observed value of $w_{\mathrm{DE}}$ at the present time. 
Therefore, at the current stage 
it seems it might not be stated that this model has an improvement 
to other dark energy models. 
However, an advantage of non-local gravity is that 
there could exist a possible scenario to screen the cosmological constant, 
although there still remains the problem of the existence of the ghost. 
This may be a clue in order to solve the smallness problem of 
the cosmological constant. 

\subsection{Cosmological consequences of adding an $R^2$ term}

We investigate the cosmological consequences of adding an $R^2$ term. 
%
%
It is known that in $F(R)$ gravity, 
the addition of an $R^2$ term could cure the finite-time future singularities 
[\cite{FS-F(R)-gravity}, in which 
the introduction of an $R^2$ term for curing the Big Rip singularity 
was first proposed, 
the first reference in 
Ref.~\cite{Review-Nojiri-Odintsov},~\cite{Future-singularity-MG, 
Singularity-F(R)}]. 
We explore whether the addition of an $R^2$ term 
removes the finite-time future singularities in non-local gravity. 
We examine the case in which 
$u R^2/\left(2\kappa^2\right)$ 
is added to the action in Eq.~(\ref{nl1}), i.e., 
the action is described by Eq.~(\ref{eq:2-F-1}). 
If the Hubble parameter is expressed as 
Eqs.~(\ref{eq:2.13}), 
in the limit $t\to t_{\mathrm{s}}$, 
$\Theta$ in Eq.~(\ref{eq:2-F-9}) and $\Xi$ in Eq.~(\ref{eq:2-F-10}) 
are approximately given by 
\begin{eqnarray} 
\hspace{-8mm}
\Theta 
\Eqn{\sim} 
18u\left[ 
-6 h_{\mathrm{s}}^2 q
\left( t_{\mathrm{s}} - t \right)^{-\left(3q+1\right)} 
+ h_{\mathrm{s}}^2 q^2 
\left( t_{\mathrm{s}} - t \right)^{-2\left(q+1\right)} 
- 2h_{\mathrm{s}}^2 q\left(q+1\right) 
\left( t_{\mathrm{s}} - t \right)^{-2\left(q+1\right)} 
\right]\,, 
\label{eq:5.12} \\
\hspace{-10mm}
\Xi \Eqn{\sim} 
6u\left[ 
9 h_{\mathrm{s}}^2 q^2
\left( t_{\mathrm{s}} - t \right)^{-2\left(q+1\right)} 
+ 18 h_{\mathrm{s}}^3 q 
\left( t_{\mathrm{s}} - t \right)^{-\left(3q+1\right)} 
\right. 
\nonumber \\ 
\hspace{-8mm}
&& \left. 
{}+ 2h_{\mathrm{s}} q\left(q+1\right)\left(q+2\right)  
\left( t_{\mathrm{s}} - t \right)^{-\left(q+3\right)} 
+ 12h_{\mathrm{s}}^2 q\left(q+1\right) 
\left( t_{\mathrm{s}} - t \right)^{-2\left(q+1\right)} 
\right]\,. 
\label{eq:5.13} 
\end{eqnarray} 
We consider the r.h.s. of Eq.~(\ref{nl7a}) with $\Theta$ 
in Eq.~(\ref{eq:2-F-9}). 
For case (ii) [$q > 1$, $\sigma < 0$],  
the first term of Eq.~(\ref{eq:5.12}), 
$-108 u h_{\mathrm{s}}^2 q 
\left( t_{\mathrm{s}} - t \right)^{-\left(3q+1\right)}$, 
becomes the leading one. 
Since $u \neq 0$, $h_{\mathrm{s}} \neq 0$ and $q \neq 0$, 
this leading term does not vanish. 
This means that 
the additional $R^2$ term can remove the finite-time future singularity. 
Furthermore, 
for case (iii) [$-1 <q < 0\,, \, 0 < q < 1$], 
the second and third terms of Eq.~(\ref{eq:5.12}), whose summation is 
given by 
$
-18u h_{\mathrm{s}}^2 q\left(q+2\right)  
\left( t_{\mathrm{s}} - t \right)^{-2\left(q+1\right)} 
$, 
becomes the leading ones. 
Since $u \neq 0$, $h_{\mathrm{s}} \neq 0$, $q \neq 0$ and 
$q \neq -2$, 
these leading terms do not vanish. 
This means that 
the additional $R^2$ term can cure the finite-time future singularity. 
The above qualitative analysis shows that adding an $R^2$ term makes 
the solutions of Secs.~II and III more complicated, admitting another 
de Sitter space. Such a de Sitter solution may be an origin of 
the inflationary stage. 
We remark that a natural non-local gravity term which may probably cure 
the finite-time future singularities 
is $R^{\upsilon} \Box^{-1} R$, 
where $\upsilon$ is a constant and 
could be close to $2$.
As this term is similar to a pure $R^2 \Box^{-1}$ operator acting on $R$, 
it cancels a number of derivative terms. 

Finally, 
it is interesting to mention that 
in the context of cosmology, the addition of an $R^2$ term might also 
be meaningful. In Ref.~\cite{Nojiri:2007uq}, 
it has been suggested that 
in the framework of 
non-local gravity combined with an $F(R)$ term, 
inflation in the early universe 
as well as the cosmic acceleration in the late time could be realized. 
The additional $R^2$ term leads to inflation 
and the late-time cosmic acceleration occurs due to 
the term of non-local gravity 
$Rf(\Box^{-1}R )$ in the action in Eq.~(\ref{nl1}). 
The unified scenario of inflation and dark energy in such a theory occurs 
qualitatively in the same way as in the models of 
$F(R)$ gravity~\cite{Review-Nojiri-Odintsov}. 

\section{Conclusion}

In this paper, we have studied 
de Sitter solutions not only in non-local gravity 
but also in non-local gravity with Lagrange constraint multiplier. 
We have explored a condition to avoid a ghost 
and presented a screening scenario for a cosmological constant 
in de Sitter solutions. 
In addition, we have explicitly shown that 
three types of the finite-time future singularities 
can occur in non-local gravity and examined their properties. 
Furthermore, 
we have investigated the behavior of the effective equation of state for 
the universe when the finite-time future singularities occur 
and shown that the late-time accelerating universe may be effectively the 
quintessence, cosmological constant or phantom-like phases. 
We have also demonstrated that there is a case with realizing 
a crossing of the phantom divide from the non-phantom (quintessence) 
phase to the phantom one in the limit of the appearance of 
a finite-time future singularity. 
The estimation of the current value of the effective equation of state 
parameter for the universe 
which could be phantomic one around -1 shows that its observed value 
could be easily realized by the appropriate choice of 
non-local gravity parameters. 
Moreover, we have considered the cosmological consequences of adding an $R^2$ 
term and demonstrated that the addition of an $R^2$ term 
can remove the finite-time future singularities in non-local gravity. 
It has also been suggested that 
the addition of an $R^2$ term in the framework of non-local gravity 
might realize unification of 
inflation in the early universe with the cosmic acceleration in the late time. 
Nevertheless, it is interesting to understand if there exists some non-local
gravity action which cures the finite-time future singularities in the similar 
way.

\section*{Acknowledgments}

We would like to thank E. O. Pozdeeva for pointing out several typos. 
We also acknowledge Ying-li Zhang for finding mistakes. 
Furthermore, we are grateful to Sergey Yu. Vernov for important suggestions.
The work is supported in part
by Global COE Program
of Nagoya University (G07) provided by the Ministry of Education, Culture,
Sports, Science \& Technology
(S.N.);
the JSPS Grant-in-Aid for Scientific Research (S) \# 22224003 (S.N.);
and
MEC (Spain) project FIS2006-02842 and AGAUR (Catalonia) 2009SGR-994
(S.D.O.).
The work of MS is supported in part  
by the Global COE Program of Kyoto University,
``The Next Generation of Physics, Spun from Universality and Emergence''
from the Ministry of Education, Culture, 
Sports, Science and Technology (MEXT) of Japan, 
by JSPS Grant-in-Aid for Scientific Research (A) No.~21244033,
by JSPS Grant-in-Aid for Creative Scientific Research No.~19GS0219, and
by Korea Institute for Advanced Study under the KIAS Scholar program.

\appendix
\section{Derivation of the leading term of $\eta$ in terms of 
$\left( t_{\mathrm{s}} - t \right)$}

In this appendix, we explain the derivation of Eqs.~(\ref{eq:2.16}) 
and (\ref{eq:2.17}) from Eq.~(\ref{eq:2.15}) in Sec.~IV A. 

For $q>1$, the integration part in terms of 
$\bar{t}$ of Eq.~(\ref{eq:2.15}) is given by 
\begin{equation} 
-\int^{\bar{t}} Ra^3 d\bar{t}
= 
\int^{\tilde{t}} 
Ra^3 d\tilde{t}
\sim 
12h_{\mathrm{s}}^2 a_{\mathrm{s}}^3 
J_1 \,,
\label{eq:A.1} 
\end{equation}
where $\tilde{t} \equiv \left( t_{\mathrm{s}} - t \right)$ 
and $J_1$ is defined as 
\begin{eqnarray} 
J_1 \Eqn{\equiv} 
\int \tilde{t}^{-2q} 
\exp \left[ \frac{3h_{\mathrm{s}}}{q-1}  
{\tilde{t}}^{-\left(q-1\right)}
\right] d\tilde{t}
\label{eq:A.2} \\ 
\Eqn{=} 
\frac{1}{\left( -3 h_{\mathrm{s}}\right)} 
\int \tilde{t}^{-2q} \tilde{t}^{q} 
\frac{d}{d \tilde{t}} 
\left\{
\exp \left[ \frac{3h_{\mathrm{s}}}{q-1}  
{\tilde{t}}^{-\left(q-1\right)}
\right] 
\right\}
d\tilde{t}
\label{eq:A.3} \\ 
\Eqn{=} 
\frac{1}{\left( -3 h_{\mathrm{s}}\right)} 
\left\{ 
\tilde{t}^{-q} 
\exp \left[ \frac{3h_{\mathrm{s}}}{q-1}  
{\tilde{t}}^{-\left(q-1\right)}
\right] 
-\left( -q \right) 
J_2 
\right\}\,, 
\label{eq:A.4}
\end{eqnarray} 
with 
\begin{eqnarray} 
J_2 
\Eqn{\equiv} 
\int 
\tilde{t}^{-\left( q+1 \right)} 
\exp \left[ \frac{3h_{\mathrm{s}}}{q-1}  
{\tilde{t}}^{-\left(q-1\right)}
\right] 
d\tilde{t} 
\label{eq:A.5} \\
\Eqn{=} 
\frac{1}{\left( -3 h_{\mathrm{s}}\right)} 
\int \tilde{t}^{-\left(q+1\right)} \tilde{t}^{q} 
\frac{d}{d \tilde{t}} 
\left\{
\exp \left[ \frac{3h_{\mathrm{s}}}{q-1} 
{\tilde{t}}^{-\left(q-1\right)}
\right] 
\right\} 
d\tilde{t}
\label{eq:A.6} \\ 
\Eqn{=} 
\frac{1}{\left( -3 h_{\mathrm{s}}\right)} 
\left\{ 
\tilde{t}^{-1} 
\exp \left[ \frac{3h_{\mathrm{s}}}{q-1}  
{\tilde{t}}^{-\left(q-1\right)}
\right] 
-\left( -1 \right) 
\int 
\tilde{t}^{-2} 
\exp \left[ \frac{3h_{\mathrm{s}}}{q-1}  
{\tilde{t}}^{-\left(q-1\right)}
\right] 
d\tilde{t}
\right\}\,. 
\label{eq:A.7} 
\end{eqnarray} 
In the second relation in (\ref{eq:A.1}), 
we have used Eq.~(\ref{eq:2.13}) and 
the fact that $\dot{H} \ll H^2$ and therefore $R \sim 12 H^2$. 
Moreover, in the equalities in Eqs.~(\ref{eq:A.4}) and (\ref{eq:A.7}), 
we have executed the partial integration. 
The leading term in terms of 
$\tilde{t} = \left( t_{\mathrm{s}} - t \right)$ in $J_1$ is 
the first term in Eqs.~(\ref{eq:A.4}). 
Using Eqs.~(\ref{eq:2.14}), (\ref{eq:A.1}) and (\ref{eq:A.4}), 
we find 
\begin{equation} 
-\int^{\bar{t}} Ra^3 d\bar{t}
= 
\int^{\tilde{t}} 
Ra^3 d\tilde{t}
\sim 
-4 h_{\mathrm{s}} 
\tilde{t}^{-q} a^3\,. 
\label{eq:A.8} 
\end{equation}
By combining Eq.~(\ref{eq:2.15}) and (\ref{eq:A.8}) 
and executing the integration with respect to $\tilde{t}$, 
we obtain Eq.~(\ref{eq:2.16}). 

On the other hand, 
for $-1 < q < 0$ and $0 < q < 1$, 
the integration part in terms of $\bar{t}$ of 
Eq.~(\ref{eq:2.15}) is given by 
\begin{equation} 
-\int^{\bar{t}} Ra^3 d \bar{t} 
= 
\int^{\tilde{t}} 
Ra^3 d\tilde{t}
\sim 
6q h_{\mathrm{s}} a_{\mathrm{s}}^3 
\int \tilde{t}^{-\left(q+1\right)} 
\exp \left[ \frac{3h_{\mathrm{s}}}{q-1}  
{\tilde{t}}^{-\left(q-1\right)}
\right] d\tilde{t}
\,, 
\label{eq:A.9} 
\end{equation}
where 
\begin{equation} 
\exp \left[ \frac{3h_{\mathrm{s}}}{q-1}  
{\tilde{t}}^{-\left(q-1\right)}
\right]
= 
1+ \frac{3h_{\mathrm{s}}}{q-1} 
{\tilde{t}}^{-\left(q-1\right)} 
+ \frac{1}{2} \left( \frac{3h_{\mathrm{s}}}{q-1} \right)^2 
{\tilde{t}}^{-2\left(q-1\right)} 
+ \dots\,. 
\label{eq:A.10}  
\end{equation}
In the second relation in Eq.~(\ref{eq:A.9}), we have used 
Eq.~(\ref{eq:2.13}) and 
the fact that $\dot{H} \gg H^2$ and hence $R \sim 6\dot{H}$. 
Since we consider the case that $-1 < q < 0$ and $0 < q < 1$, 
in the limit of $t \to t_{\mathrm{s}}$, i.e., $\tilde{t} \to 0$, 
from Eq.~(\ref{eq:A.10}) 
we see that 
the first term on the r.h.s. of Eq.~(\ref{eq:A.10}) is the leading one. 
Hence, we have 
\begin{equation} 
-\int^{t^{\prime}} Ra^3 dt^{\prime}
= 
\int^{\tilde{t}} 
Ra^3 d\tilde{t}
\sim 
-6 h_{\mathrm{s}} a_{\mathrm{s}}^3  
{\tilde{t}}^{-q}\,. 
\label{eq:A.11} 
\end{equation}
Substituting Eq.~(\ref{eq:A.11}) into Eq.~(\ref{eq:2.15}) 
and using Eq.~(\ref{eq:2.14}), 
we find 
\begin{equation} 
\eta = 6 h_{\mathrm{s}} \int 
\tilde{t}^{-q} 
\exp \left[ -\frac{3h_{\mathrm{s}}}{q-1}  
{\tilde{t}}^{-\left(q-1\right)}
\right] d\tilde{t}\,, 
\label{eq:A.12} 
\end{equation}
where 
\begin{equation} 
\exp \left[ -\frac{3h_{\mathrm{s}}}{q-1}  
{\tilde{t}}^{-\left(q-1\right)}
\right]
= 
1 - \frac{3h_{\mathrm{s}}}{q-1} 
{\tilde{t}}^{-\left(q-1\right)} 
+ \frac{1}{2} \left( -\frac{3h_{\mathrm{s}}}{q-1} \right)^2 
{\tilde{t}}^{-2\left(q-1\right)} 
+ \dots\,. 
\label{eq:A.13} 
\end{equation}
Similarly to Eq.~(\ref{eq:A.10}), 
the first term on the r.h.s. of Eq.~(\ref{eq:A.13}) is the leading one. 
Thus, by executing the integration with respect to $\tilde{t}$ 
in Eq.~(\ref{eq:A.12}) and taking the leading term in terms of 
$\tilde{t} = \left( t_{\mathrm{s}} - t \right)$, 
we acquire Eq.~(\ref{eq:2.17}).



\begin{thebibliography}{99}

\bibitem{SN1}
%
 S.~Perlmutter {\it et al.}  [SNCP Collaboration],
 Astrophys.\ J.\  {\bf 517}, 565 (1999) 
 [arXiv:astro-ph/9812133]; \\ 
%
  A.~G.~Riess {\it et al.}  [Supernova Search Team Collaboration],
  Astron.\ J.\  {\bf 116}, 1009 (1998)
  [arXiv:astro-ph/9805201].
%

\bibitem{WMAP}
%
D.~N.~Spergel {\it et al.}  [WMAP Collaboration],
Astrophys.\ J.\ Suppl.\  {\bf 148}, 175 (2003) 
[arXiv:astro-ph/0302209]; \\ 
%
 D.~N.~Spergel {\it et al.}  [WMAP Collaboration],
 Astrophys.\ J.\ Suppl.\  {\bf 170}, 377 (2007) 
 [arXiv:astro-ph/0603449]; \\
%
 E.~Komatsu {\it et al.}  [WMAP Collaboration],
 Astrophys.\ J.\ Suppl.\  {\bf 180}, 330 (2009) 
 [arXiv:0803.0547 [astro-ph]]. 
%

\bibitem{Komatsu:2010fb}
E.~Komatsu {\it et al.}  [WMAP Collaboration],
Astrophys.\ J.\ Suppl.\ {\bf 192}, 18 (2011)
[arXiv:1001.4538 [astro-ph.CO]].

\bibitem{LSS}
%
  M.~Tegmark {\it et al.}  [SDSS Collaboration],
  Phys.\ Rev.\  D {\bf 69}, 103501 (2004)
  [arXiv:astro-ph/0310723]; \\
%
  U.~Seljak {\it et al.}  [SDSS Collaboration],
  Phys.\ Rev.\  D {\bf 71}, 103515 (2005)
  [arXiv:astro-ph/0407372].
%

\bibitem{Eisenstein:2005su}
  D.~J.~Eisenstein {\it et al.}  [SDSS Collaboration],
  Astrophys.\ J.\  {\bf 633}, 560 (2005)
  [arXiv:astro-ph/0501171].

\bibitem{Jain:2003tba}
  B.~Jain and A.~Taylor,
  Phys.\ Rev.\ Lett.\  {\bf 91}, 141302 (2003)
  [arXiv:astro-ph/0306046].

\bibitem{Weinberg:1988cp}
  S.~Weinberg,
  Rev.\ Mod.\ Phys.\  {\bf 61}, 1 (1989).


\bibitem{Review-Nojiri-Odintsov}
%
  S.~Nojiri and S.~D.~Odintsov,
  Phys.\ Rept.\  {\bf 505}, 59 (2011)
  [arXiv:1011.0544 [gr-qc]]; \\
%
 S.~Nojiri and S.~D.~Odintsov,
 eConf {\bf C0602061}, 06 (2006)
 [Int.\ J.\ Geom.\ Meth.\ Mod.\ Phys.\  {\bf 4}, 115 (2007)]
 [arXiv:hep-th/0601213].

\bibitem{Book-Capozziello-Faraoni}
S.~Capozziello and V.~Faraoni,
\textit{Beyond Einstein Gravity}
(Springer, 2010). 

\bibitem{Deser:2007jk}
  S.~Deser and R.~P.~Woodard,
  Phys.\ Rev.\ Lett.\  {\bf 99}, 111301 (2007)
  [arXiv:0706.2151 [astro-ph]].

\bibitem{Nojiri:2007uq}
  S.~Nojiri and S.~D.~Odintsov,
  Phys.\ Lett.\  B {\bf 659}, 821 (2008)
  [arXiv:0708.0924 [hep-th]].

\bibitem{ArkaniHamed:2002fu}
  N.~Arkani-Hamed, S.~Dimopoulos, G.~Dvali and G.~Gabadadze,
  arXiv:hep-th/0209227.

\bibitem{Nojiri:2010pw}
  S.~Nojiri, S.~D.~Odintsov, M.~Sasaki and Y.~l.~Zhang,
  Phys.\ Lett.\  B {\bf 696}, 278 (2011)
  [arXiv:1010.5375 [gr-qc]].
 

\bibitem{Non-local-gravity-Refs}
%
  L.~Joukovskaya,
  Phys.\ Rev.\  D {\bf 76}, 105007 (2007)
  [arXiv:0707.1545 [hep-th]]; \\
%
  G.~Calcagni, M.~Montobbio and G.~Nardelli,
  Phys.\ Lett.\  B {\bf 662}, 285 (2008)
  [arXiv:0712.2237 [hep-th]]; \\
%
  S.~Jhingan, S.~Nojiri, S.~D.~Odintsov, M.~Sami, I.~Thongkool and S.~Zerbini,
  Phys.\ Lett.\  B {\bf 663}, 424 (2008)
  [arXiv:0803.2613 [hep-th]]; \\
%
  S.~Capozziello, E.~Elizalde, S.~Nojiri and S.~D.~Odintsov,
  Phys.\ Lett.\  B {\bf 671}, 193 (2009)
  [arXiv:0809.1535 [hep-th]]; \\
%
  N.~A.~Koshelev,
  Grav.\ Cosmol.\  {\bf 15}, 220 (2009)
  [arXiv:0809.4927 [gr-qc]]; \\
%
  S.~Nesseris and A.~Mazumdar,
  Phys.\ Rev.\  D {\bf 79}, 104006 (2009)
  [arXiv:0902.1185 [astro-ph.CO]]; \\
%
  C.~Deffayet and R.~P.~Woodard,
  JCAP {\bf 0908}, 023 (2009)
  [arXiv:0904.0961 [gr-qc]]; \\
%
  G.~Calcagni and G.~Nardelli,
  Int.\ J.\ Mod.\ Phys.\  D {\bf 19}, 329 (2010)
  [arXiv:0904.4245 [hep-th]]; \\
%
  G.~Cognola, E.~Elizalde, S.~Nojiri, S.~D.~Odintsov and S.~Zerbini,
  Eur.\ Phys.\ J.\  C {\bf 64}, 483 (2009)
  [arXiv:0905.0543 [gr-qc]]; \\
%
  K.~A.~Bronnikov and E.~Elizalde,
  Phys.\ Rev.\  D {\bf 81}, 044032 (2010)
  [arXiv:0910.3929 [hep-th]]; \\
%
  G.~Calcagni and G.~Nardelli,
  Phys.\ Rev.\  D {\bf 82}, 123518 (2010)
  [arXiv:1004.5144 [hep-th]]; \\
%
  S.~Y.~Vernov,
  arXiv:1005.0372 [astro-ph.CO]; \\
%
  T.~Biswas, T.~Koivisto and A.~Mazumdar,
  JCAP {\bf 1011}, 008 (2010)
  [arXiv:1005.0590 [hep-th]]; \\ 
%
  N.~Barnaby,
  Nucl.\ Phys.\  B {\bf 845}, 1 (2011)
  [arXiv:1005.2945 [hep-th]].
%

%
\bibitem{Koivisto:2008xfa}
  T.~Koivisto,
  Phys.\ Rev.\  D {\bf 77}, 123513 (2008)
  [arXiv:0803.3399 [gr-qc]].
%

\bibitem{Koivisto:2008dh}
  T.~S.~Koivisto,
  Phys.\ Rev.\  D {\bf 78}, 123505 (2008)
  [arXiv:0807.3778 [gr-qc]].


\bibitem{Elizalde:2011su}
  E.~Elizalde, E.~O.~Pozdeeva and S.~Y.~Vernov,
  arXiv:1110.5806 [astro-ph.CO].


\bibitem{Dolgov:2003px}
  A.~D.~Dolgov and M.~Kawasaki,
  Phys.\ Lett.\  B {\bf 573}, 1 (2003)
  [arXiv:astro-ph/0307285].

\bibitem{maeda}
%
  T.~Kobayashi and K.~i.~Maeda,
  Phys.\ Rev.\  D {\bf 78}, 064019 (2008)
  [arXiv:0807.2503 [astro-ph]]; \\ 
%
  T.~Kobayashi and K.~i.~Maeda,
  Phys.\ Rev.\  D {\bf 79}, 024009 (2009)
  [arXiv:0810.5664 [astro-ph]]; \\
%
  A.~Dev, D.~Jain, S.~Jhingan, S.~Nojiri, M.~Sami and I.~Thongkool,
  Phys.\ Rev.\  D {\bf 78}, 083515 (2008)
  [arXiv:0807.3445 [hep-th]]. 
%

\bibitem{Arbuzova:2010iu}
  E.~V.~Arbuzova and A.~D.~Dolgov,
  Phys.\ Lett.\  B {\bf 700}, 289 (2011)
  [arXiv:1012.1963 [astro-ph.CO]].

\bibitem{Bamba:2011sm}
  K.~Bamba, S.~Nojiri and S.~D.~Odintsov,
  Phys.\ Lett.\  B {\bf 698}, 451 (2011)
  [arXiv:1101.2820 [gr-qc]].

\bibitem{Conformal-anomalies}
%
  S.~Deser,
  Phys.\ Lett.\  B {\bf 479}, 315 (2000)
  [arXiv:hep-th/9911129];\ 
%
  S.~Deser,
  Nucl.\ Phys.\ Proc.\ Suppl.\  {\bf 88}, 204 (2000).
%

\bibitem{Deruelle:2010kf}
  N.~Deruelle, M.~Sasaki, Y.~Sendouda and A.~Youssef,
  JCAP {\bf 1103}, 040 (2011)
  [arXiv:1012.5202 [gr-qc]].

\bibitem{withghost}
  S.~W.~Hawking and T.~Hertog,
  Phys.\ Rev.\  D {\bf 65}, 103515 (2002)
  [arXiv:hep-th/0107088]; \\
  T.~Clunan and M.~Sasaki,
  Class.\ Quant.\ Grav.\  {\bf 27}, 165014 (2010)
  [arXiv:0907.3868 [hep-th]]; \\
  M.~Fontanini and M.~Trodden,
  Phys.\ Rev.\  D {\bf 83}, 103518 (2011)
  [arXiv:1102.4357 [gr-qc]].


\bibitem{Biswas:2005vz}
  T.~Biswas and A.~Notari,
  Phys.\ Rev.\  D {\bf 74}, 043508 (2006)
  [arXiv:hep-ph/0511207].

%
%
\bibitem{Zhang:2011uv}
  Y.~l.~Zhang and M.~Sasaki,
  Int.\ J.\ Mod.\ Phys.\  D {\bf 21}, 1250006 (2012)
  [arXiv:1108.2112 [gr-qc]].

\bibitem{Nojiri:2009th}
  S.~Nojiri and S.~D.~Odintsov,
  Phys.\ Rev.\  D {\bf 81}, 043001 (2010)
  [arXiv:0905.4213 [hep-th]].

\bibitem{Nojiri:2010tv}
  S.~Nojiri and S.~D.~Odintsov,
  Phys.\ Lett.\  B {\bf 691}, 60 (2010)
  [arXiv:1004.3613 [hep-th]].

\bibitem{Horava:2009uw}
  P.~Horava,
  Phys.\ Rev.\  D {\bf 79}, 084008 (2009)
  [arXiv:0901.3775 [hep-th]].

\bibitem{Lagrange-Constraint-Multiplier}
%
  E.~A.~Lim, I.~Sawicki and A.~Vikman,
  JCAP {\bf 1005}, 012 (2010)
  [arXiv:1003.5751 [astro-ph.CO]]; \\
%
  C.~Gao, Y.~Gong, X.~Wang and X.~Chen,
  Phys.\ Lett.\  B {\bf 702}, 107 (2011)
  [arXiv:1003.6056 [astro-ph.CO]].
%

\bibitem{FS-F(R)-gravity}
%
  M.~C.~B.~Abdalla, S.~Nojiri and S.~D.~Odintsov,
  Class.\ Quant.\ Grav.\  {\bf 22}, L35 (2005)
  [arXiv:hep-th/0409177]; \\
%
  F.~Briscese, E.~Elizalde, S.~Nojiri and S.~D.~Odintsov,
  Phys.\ Lett.\  B {\bf 646}, 105 (2007)
  [arXiv:hep-th/0612220].
%

\bibitem{Future-singularity-MG}
%
  K.~Bamba, S.~Nojiri and S.~D.~Odintsov,
  JCAP {\bf 0810}, 045 (2008)
  [arXiv:0807.2575 [hep-th]]; \\
%
  K.~Bamba, S.~D.~Odintsov, L.~Sebastiani and S.~Zerbini,
  Eur.\ Phys.\ J.\  C {\bf 67}, 295 (2010)
  [arXiv:0911.4390 [hep-th]].
%

\bibitem{Nojiri:2005sx}
  S.~Nojiri, S.~D.~Odintsov and S.~Tsujikawa,
  Phys.\ Rev.\  D {\bf 71}, 063004 (2005)
  [arXiv:hep-th/0501025].

\bibitem{Big-Rip}
%
  R.~R.~Caldwell, M.~Kamionkowski and N.~N.~Weinberg,
  Phys.\ Rev.\ Lett.\  {\bf 91}, 071301 (2003)
  [arXiv:astro-ph/0302506]; \\
%
B.~McInnes,
JHEP {\bf 0208} (2002) 029 
[arXiv:hep-th/0112066]; \\
%
S.~Nojiri and S.~D.~Odintsov,
Phys.\ Lett.\ B {\bf 562}, 147 (2003)
[arXiv:hep-th/0303117]; \\
%
S.~Nojiri and S.~D.~Odintsov,
Phys.\ Lett.\ B {\bf 571}, 1 (2003)
[arXiv:hep-th/0306212]; \\
%
V.~Faraoni,
Int.\ J.\ Mod.\ Phys.\ D {\bf 11}, 471 (2002)
[arXiv:astro-ph/0110067]; \\
%
P.~F.~Gonzalez-Diaz,
Phys.\ Lett.\ B {\bf 586}, 1 (2004)
[arXiv:astro-ph/0312579]; \\
%
  E.~Elizalde, S.~Nojiri and S.~D.~Odintsov,
  Phys.\ Rev.\  D {\bf 70}, 043539 (2004)
  [arXiv:hep-th/0405034]; \\
%
B.~McInnes,
Nucl.\ Phys.\ B {\bf 718}, 55 (2005)
[arXiv:hep-th/0502209]; \\
%
P.~Singh, M.~Sami and N.~Dadhich,
Phys.\ Rev.\ D {\bf 68}, 023522 (2003)
[arXiv:hep-th/0305110]; \\
%
C.~Csaki, N.~Kaloper and J.~Terning,
Annals Phys.\ {\bf 317}, 410 (2005)
[arXiv:astro-ph/0409596]; \\
%
P.~X.~Wu and H.~W.~Yu,
Nucl.\ Phys.\ B {\bf 727}, 355 (2005)
[arXiv:astro-ph/0407424]; \\
%
S.~Nesseris and L.~Perivolaropoulos,
Phys.\ Rev.\ D {\bf 70}, 123529 (2004)
[arXiv:astro-ph/0410309]; \\
%
M.~Sami and A.~Toporensky,
Mod.\ Phys.\ Lett.\ A {\bf 19}, 1509 (2004)
[arXiv:gr-qc/0312009]; \\
%
H.~Stefancic,
Phys.\ Lett.\ B {\bf 586}, 5 (2004)
[arXiv:astro-ph/0310904]; \\
%
L.~P.~Chimento and R.~Lazkoz,
Phys.\ Rev.\ Lett.\ {\bf 91}, 211301 (2003)
[arXiv:gr-qc/0307111]; \\
%
J.~G.~Hao and X.~Z.~Li,
Phys.\ Lett.\ B {\bf 606}, 7 (2005)
[arXiv:astro-ph/0404154]; \\
%
E.~Babichev, V.~Dokuchaev and Yu.~Eroshenko,
Class.\ Quant.\ Grav.\ {\bf 22}, 143 (2005)
[arXiv:astro-ph/0407190]; \\
%
X.~F.~Zhang, H.~Li, Y.~S.~Piao and X.~M.~Zhang,
Mod.\ Phys.\ Lett.\ A {\bf 21}, 231 (2006)
[arXiv:astro-ph/0501652]; \\
%
E.~Elizalde, S.~Nojiri, S.~D.~Odintsov and P.~Wang,
Phys.\ Rev.\ D {\bf 71}, 103504 (2005)
[arXiv:hep-th/0502082]; \\
%
M.~P.~Dabrowski and T.~Stachowiak,
Annals Phys.\ {\bf 321}, 771 (2006)
[arXiv:hep-th/0411199]; \\
%
F.~S.~N.~Lobo,
Phys.\ Rev.\ D {\bf 71}, 084011 (2005)
[arXiv:gr-qc/0502099]; \\
%
R.~G.~Cai, H.~S.~Zhang and A.~Wang,
Commun.\ Theor.\ Phys.\ {\bf 44}, 948 (2005)
[arXiv:hep-th/0505186]; \\
%
  I.~Y.~Aref'eva, A.~S.~Koshelev and S.~Y.~Vernov,
  Phys.\ Rev.\  D {\bf 72}, 064017 (2005)
  [arXiv:astro-ph/0507067]; \\
%
H.~Q.~Lu, Z.~G.~Huang and W.~Fang,
arXiv:hep-th/0504038; \\
%
W.~Godlowski and M.~Szydlowski,
Phys.\ Lett.\ B {\bf 623}, 10 (2005)
[arXiv:astro-ph/0507322]; \\
%
J.~Sola and H.~Stefancic,
Phys.\ Lett.\ B {\bf 624}, 147 (2005)
[arXiv:astro-ph/0505133]; \\
%
B.~Guberina, R.~Horvat and H.~Nikolic,
Phys.\ Rev.\ D {\bf 72}, 125011 (2005)
[arXiv:astro-ph/0507666]; \\
%
M.~P.~Dabrowski, C.~Kiefer and B.~Sandhofer,
Phys.\ Rev.\ D {\bf 74}, 044022 (2006)
[arXiv:hep-th/0605229]; \\
%
  E.~M.~.~Barbaoza and N.~A.~Lemos,
  arXiv:gr-qc/0606084.
%

\bibitem{Shtanov:2002ek}
  Y.~Shtanov and V.~Sahni,
  Class.\ Quant.\ Grav.\  {\bf 19}, L101 (2002)
  [arXiv:gr-qc/0204040].

\bibitem{sudden}
%
 J.~D.~Barrow,
  Class.\ Quant.\ Grav.\  {\bf 21}, L79 (2004)
  [arXiv:gr-qc/0403084]; \\
%
 S.~Nojiri and S.~D.~Odintsov,
 Phys.\ Lett.\  B {\bf 595}, 1 (2004)
  [arXiv:hep-th/0405078]; \\
%
 S.~Nojiri and S.~D.~Odintsov,
 Phys.\ Rev.\  D {\bf 70}, 103522 (2004)
  [arXiv:hep-th/0408170]; \\
%
 S.~Cotsakis and I.~Klaoudatou,
 J.\ Geom.\ Phys.\  {\bf 55}, 306 (2005) 
  [arXiv:gr-qc/0409022]; \\
%
 M.~P.~Dabrowski,
 Phys.\ Rev.\  D {\bf 71}, 103505 (2005) 
  [arXiv:gr-qc/0410033]; \\
%
 L.~Fernandez-Jambrina and R.~Lazkoz,
 Phys.\ Rev.\  D {\bf 70}, 121503 (2004) 
  [arXiv:gr-qc/0410124]; \\
%
 J.~D.~Barrow and C.~G.~Tsagas,
 Class.\ Quant.\ Grav.\  {\bf 22}, 1563 (2005) 
  [arXiv:gr-qc/0411045]; \\
%
 H.~Stefancic,
 Phys.\ Rev.\  D {\bf 71}, 084024 (2005)
  [arXiv:astro-ph/0411630]; \\
%
  S.~Nojiri and S.~D.~Odintsov,
  Phys.\ Rev.\  D {\bf 72}, 023003 (2005)
  [arXiv:hep-th/0505215]; \\
%
 M.~P.~Dabrowski,
 Phys.\ Lett.\  B {\bf 625}, 184 (2005) 
  [arXiv:gr-qc/0505069]; \\ 
%
 C.~Cattoen and M.~Visser,
 Class.\ Quant.\ Grav.\  {\bf 22}, 4913 (2005) 
  [arXiv:gr-qc/0508045]; \\
%
 P.~Tretyakov, A.~Toporensky, Y.~Shtanov and V.~Sahni,
 Class.\ Quant.\ Grav.\  {\bf 23}, 3259 (2006) 
  [arXiv:gr-qc/0510104]; \\
%
 A.~Balcerzak and M.~P.~Dabrowski,
 Phys.\ Rev.\  D {\bf 73}, 101301 (2006) 
  [arXiv:hep-th/0604034]; \\
%
 M.~Sami, P.~Singh and S.~Tsujikawa,
  Phys.\ Rev.\  D {\bf 74}, 043514 (2006) 
  [arXiv:gr-qc/0605113]; \\
%
 M.~Bouhmadi-Lopez, P.~F.~Gonzalez-Diaz and P.~Martin-Moruno,
 Phys.\ Lett.\  B {\bf 659}, 1 (2008)
  [arXiv:gr-qc/0612135]; \\
%
 A.~V.~Yurov, A.~V.~Astashenok and P.~F.~Gonzalez-Diaz,
 Grav.\ Cosmol.\  {\bf 14}, 205 (2008) 
  [arXiv:0705.4108 [astro-ph]]; \\
%
  L.~Fernandez-Jambrina and R.~Lazkoz,
 Phys.\ Lett.\  B {\bf 670}, 254 (2009)
  [arXiv:0805.2284 [gr-qc]]; \\
%
 I.~Brevik and O.~Gorbunova,
 Eur.\ Phys.\ J.\  C {\bf 56}, 425 (2008)
  [arXiv:0806.1399 [gr-qc]]; \\
%
 J.~D.~Barrow and S.~Z.~W.~Lip,
 Phys.\ Rev.\  D {\bf 80}, 043518 (2009) 
  [arXiv:0901.1626 [gr-qc]]; \\
%
  M.~Bouhmadi-Lopez, Y.~Tavakoli and P.~V.~Moniz,
  JCAP {\bf 1004}, 016 (2010)
  [arXiv:0911.1428 [gr-qc]].

\bibitem{observational status}
%
  U.~Alam, V.~Sahni and A.~A.~Starobinsky,
  JCAP {\bf 0406}, 008 (2004) 
  [arXiv:astro-ph/0403687]; \\
%
  S.~Nesseris and L.~Perivolaropoulos, 
  JCAP {\bf 0701}, 018 (2007)
  [arXiv:astro-ph/0610092]; \\
%
  P.~U.~Wu and H.~W.~Yu,
  Phys.\ Lett.\  B {\bf 643}, 315 (2006)
  [arXiv:astro-ph/0611507]; \\
%
  H.~K.~Jassal, J.~S.~Bagla and T.~Padmanabhan,
  Mon.\ Not.\ Roy.\ Astron.\ Soc.\  {\bf 405}, 2639 (2010) 
  [arXiv:astro-ph/0601389]. 
%

\bibitem{Future-Crossing-in-F(R)}
%
  K.~Bamba, C.~Q.~Geng, S.~Nojiri and S.~D.~Odintsov,
  Phys.\ Rev.\  D {\bf 79}, 083014 (2009)
  [arXiv:0810.4296 [hep-th]]; \\
%
  K.~Bamba, C.~Q.~Geng and C.~C.~Lee,
  JCAP {\bf 1011}, 001 (2010)
  [arXiv:1007.0482 [astro-ph.CO]]. 
%

\bibitem{C-P-L}
%
  M.~Chevallier and D.~Polarski,
  Int.\ J.\ Mod.\ Phys.\  D {\bf 10}, 213 (2001)
  [arXiv:gr-qc/0009008]; \\
%
  E.~V.~Linder,
  Phys.\ Rev.\ Lett.\  {\bf 90}, 091301 (2003)
  [arXiv:astro-ph/0208512].

\bibitem{Kolb and Turner}
E.~W.~Kolb and M.~S.~Turner,
\textit{The Early Universe}
(Addison-Wesley, Redwood City, California, 1990).

\bibitem{Freedman:2000cf}
 W.~L.~Freedman {\it et al.}  [HST Collaboration],
 Astrophys.\ J.\  {\bf 553}, 47 (2001) 
 [arXiv:astro-ph/0012376].

\bibitem{Singularity-F(R)}
%
 S.~Nojiri and S.~D.~Odintsov,
 Phys.\ Rev.\  D {\bf 78}, 046006 (2008)
 [arXiv:0804.3519 [hep-th]]; \\
%
 S.~Capozziello, M.~De Laurentis, S.~Nojiri and S.~D.~Odintsov,
 Phys.\ Rev.\  D {\bf 79}, 124007 (2009)
 [arXiv:0903.2753 [hep-th]].
%

\end{thebibliography}
\end{document}